\documentclass[lettersize,journal]{IEEEtran}
\usepackage{amsmath,amsfonts}
\usepackage{amsthm}
\usepackage{algorithmic}
\usepackage{multirow}
\usepackage{booktabs}
\usepackage[ruled,linesnumbered]{algorithm2e}
\usepackage{array}
\usepackage[caption=false,font=normalsize,labelfont=sf,textfont=sf]{subfig}
\usepackage{textcomp}
\usepackage{stfloats}
\usepackage{url}
\usepackage{verbatim}
\usepackage{graphicx}
\usepackage[colorlinks,citecolor=blue,linkcolor=blue,urlcolor=blue]{hyperref}
\usepackage{cite}
\usepackage{cancel}
\usepackage{indentfirst}
\usepackage{cite}
\usepackage{color}
\usepackage{mdframed}

\begin{document}
	
	\title{Hybrid Architecture Gets Fluid: A New Paradigm for Direction-of-arrival Estimation in 6G Networks}
%\title{Hybrid Analog-Digital Fluid Antenna Systems: Design and Direction-of-Arrival Estimation}
\author{Ye Tian, Jiaji Ren, Tuo Wu, Wei Liu, Maged Elkashlan,   Matthew C. Valenti, \emph{Fellow}, \emph{IEEE}, \\ Naofal Al-Dhahir, \emph{Fellow, IEEE},   and  Hing Cheung So, \emph{Fellow, IEEE}
	\thanks{   \emph{(Corresponding author: Tuo Wu.)}
		
		Y. Tian and J. Ren are with the Faculty of Electrical Engineering and Computer Science, Ningbo University, Ningbo 315211, China (E-mail: $\rm tianfield@126.com; 2311100053@nbu.edu.cn$).
		T. Wu and H. C. So are with the Department of Electrical Engineering City University of Hong Kong, Hong Kong, China. (E-mail: $\rm tuo.wu@qmul.ac.uk; hcso@ee.cityu.edu.hk$).
		W. Liu is with the Department of Electrical and Electronic Engineering, Hong Kong Polytechnic University, Kowloon, Hong Kong, China (E-mail: $\rm wei2.liu@polyu.edu.hk$).
		M. Elkashlan is with the School of Electronic Engineering and Computer Science at Queen Mary University of London, London E1 4NS, U.K. (E-mail: $\rm maged.elkashlan@qmul.ac.uk$). 
		M. Valenti is with the Lane Department of Computer Science and Electrical Engineering, West Virginia University, Morgantown, USA (E-mail: $\rm valenti@ieee.org$).  Naofal Al-Dhahir is with the Department of Electrical and Computer Engineering, The University of Texas at Dallas, Richardson, TX 75080 USA (E-mail: $\rm aldhahir@utdallas.edu$).
	}
}
	%\markboth{Journal of \LaTeX\ Class Files,~Vol.~XX, No.~XX, XX~XXXX}%

	
	\maketitle	
	\begin{abstract}
		High-precision direction-of-arrival (DOA) estimation, as a key sensing capability for 6G-enabled applications such as autonomous driving and extended reality, is increasingly dependent on the effective exploitation of spatial degrees of freedom (DOFs). This paper integrates two frontier DOFs-oriented paradigms and proposes a fluid antenna-enabled hybrid analog-digital (FA-HAD) architecture, which features an extremely lightweight front-end configuration mechanism and efficient spatial DOFs exploitation. Within this architecture, a collaborative spatial-phase sampling strategy is first developed to enable real-time 2-D DOA estimation under compressive observations, and a single-source CRLB analysis is provided to quantify the achievable performance limit, offering quantitative guidance for accuracy-overhead trade-offs. Furthermore, an efficient virtual-array spatial covariance matrix  reconstruction method is proposed to recover a physically meaningful covariance representation, thereby providing a covariance-domain interface that is directly reusable by a broad class of existing covariance-based array processing and array design techniques, which strengthens the scalability and transferability of the proposed architecture. Building upon the reconstructed SCM, a Jacobi-Anger expansion based dimension-reduced MUSIC estimator is further derived for arbitrary planar arrays with a favorable computational cost. Simulation results demonstrate that the proposed FA-HAD framework attains DOA accuracy close to fully digital systems while substantially reducing RF hardware complexity and training overhead.
	\end{abstract}
	\section{Introduction}
	
	\IEEEPARstart{A}{s} the cornerstone of future communication technologies, the sixth-generation (6G) mobile communication network is poised to enable a wide array of revolutionary applications, including interactive immersive experiences, holographic communication, fully autonomous driving, real-time intelligent healthcare, and collaborative unmanned factories \cite{r1},\cite{r2},\cite{r3}. In this transformative landscape, direction-of-arrival (DOA) estimation, as a critical sensing technology, will play an indispensable role in emerging applications such as autonomous driving networks, indoor positioning and activity recognition, unmanned aerial vehicle (UAV) communication and sensing, extended reality (XR), and integrated radar-communication systems, providing essential technical support for intelligent and ubiquitous communication and sensing \cite{r4},\cite{r5},\cite{r6}. Deep exploitation of spatial degrees of freedom (DOFs) has always been a key enabler for achieving high-precision DOA estimation: richer spatial DOFs allow a larger effective aperture and a higher-dimensional spatial sampling space, thereby yielding higher spatial resolution, stronger interference suppression capability, and more robust multi-target resolvability. Although DOA estimation based on fixed fully-digital (FD) architectures has been extensively studied, its strong dependence on the number of radio-frequency (RF) chains, power consumption, and hardware scale hinders the continued expansion of spatial DOFs, making it difficult to simultaneously achieve cost-efficiency and high performance in large-scale 6G deployments. Therefore, exploring new system paradigms with better scalability and cost efficiency becomes a natural and inevitable choice.
	
	Recent advances beyond the conventional fixed FD array paradigm can be broadly categorized into two representative directions. The first direction is the hybrid analog--digital (HAD) architecture \cite{r15},\cite{r16},\cite{r17},\cite{r18}, whose primary goal is to reduce deployment cost, thereby enabling further array scaling and unlocking richer DOFs. Compared with the conventional fixed FD architecture, HAD typically adopts a modular design by partitioning a large-scale array into subarrays and, with highly integrated and low-cost analog beamformers, establishes scalable connection and selection mechanisms between RF chains and subarrays, thereby substantially reducing the required number of  RF chains\cite{r19},\cite{r20}. Consequently, HAD effectively alleviates hardware cost and power consumption burdens, and improves practical deployability in large-scale multiple-input multiple-output (MIMO) scenarios.
	
	The second direction is the fluid antenna system (FAS), which leverages reconfigurable hardware mechanisms to provide transformative capabilities for wireless localization and sensing \cite{r50, r21,r22,r23,r24,r51, r52}.  Distinct from the fixed FD paradigm, FAS introduces a new ``spatially reconfigurable'' paradigm  \cite{r26}: within a constrained deployment region, antenna positions can be reconfigured and selected over a continuum of spatial locations, enabling more thorough exploitation and utilization of spatial DOFs in a limited space. Moreover, by deploying a small number of fluid antennas and performing multiple spatial samplings, FAS can emulate the functionality of large-scale arrays without a proportional increase in hardware, thereby achieving effective aperture expansion and performance enhancement \cite{r25}.

	Existing studies have confirmed the remarkable performance advantages of fluid antenna (FA) technology in wireless communications \cite{r27,r28, r29,r30,r31,r32,r33,r34,r35,r36,r44,r43}. The superiority of a FAS in multi-access scenarios has been systematically demonstrated in \cite{r27,r28, r29} through theoretical proofs and numerical simulations, while \cite{r30} employs analytical methods to model the outage probability and diversity gain, revealing FAS's potential for high-efficiency communication under low-complexity configurations.  {Beyond multi-access, the FAS paradigm has been extended to cooperative non-orthogonal multiple access (CoNOMA) systems \cite{r31}, intelligent antenna positioning for integrated sensing and communication (ISAC) via deep reinforcement learning \cite{r32}, and vehicle-to-everything (V2X) communications in conjunction with reconfigurable intelligent surfaces (RIS) \cite{r33}, collectively demonstrating the broad applicability of FAS across diverse scenarios. Furthermore, significant advancements in channel modeling have been achieved: \cite{r34} revisits the spatial block-correlation model by generalizing from constant to variable correlations, while \cite{r35} further develops variable block-correlation modeling and optimization for secrecy analysis.} Additionally, compressed sensing-based approaches in \cite{r36} enable high-accuracy joint estimation of multipath parameters, establishing new paradigms for dynamic environment communications. More recently, \cite{r44} investigates DOA estimation under sparse fluid antenna configurations, where a sparse-motion spatial sampling paradigm is developed to achieve strong estimation performance with only a small number of carefully designed array movements, providing useful insights into the interplay between mobility cost and estimation accuracy.
	
	Although FAS have demonstrated unique advantages in wireless communications and sensing, to the best of our knowledge, there is still a lack of systematic studies that integrate FAS into HAD architectures and evaluate their potential gains and key challenges for DOA estimation. Moreover, the practical deployment of FAS is often constrained by factors such as mobility cost, the increased number of measurements and pilot overhead, and the complexity of multi-antenna trajectory scheduling. Under these system constraints, combining FAS with HAD should be viewed as a complementary synergy rather than a simple superposition of their individual benefits: FAS provides reconfigurability at the geometry and spatial-sampling levels, endowing the system with enhanced adaptivity and richer capability to exploit spatial DOFs; HAD, in turn, offers an extremely lightweight front-end configuration mechanism together with a scalable observation and processing framework, thereby supporting large-scale deployments with low hardware overhead. More importantly, the inherent modular signal aggregation and processing structure in HAD (i.e., the ``RF chain--to--subarray'' mapping) provides an engineering interface for multi-position sampling in FAS, making modular mobility and sampling a more implementable design trade-off.

   However, realizing the integration of FAS and HAD architectures for practical DOA estimation still requires systematically addressing three key issues. First, the analog-domain compressive observation nature of HAD makes it considerably more challenging to achieve reliable DOA estimation and to establish rigorous theoretical analysis; whether such a fusion can deliver substantial performance gains therefore remains to be answered. Second, it is yet to be verified whether the fused architecture can retain good scalability and naturally interface with the extensive existing studies on FAS and DOA estimation, so as to support engineering evolution toward larger-scale and more complex configurations. Finally, the flexible configuration of FAS often leads to irregular virtual-array geometries, which violates the structured-array assumptions underlying most high-resolution DOA estimators; meanwhile, wide-angle sensing for planar arrays can further raise the computational burden significantly. Therefore, it is imperative to develop new DOA estimation frameworks that are both broadly applicable and computationally affordable.

	Motivated by the above challenges, this paper proposes a fluid antenna-enabled HAD (FA-HAD) architecture, together with a systematic DOA estimation framework tailored for low-overhead deployment. Specifically, we design a coordinated spatial-phase sampling mechanism to support real-time angle estimation, and characterize the achievable performance limits under compressive observations via single-source Cram\'{e}r-Rao Lower Bound (CRLB) derivations. We further develop an efficient virtual-array spatial covariance matrix (SCM) reconstruction method, recovering a covariance-domain representation that enables compatible interfacing with existing covariance-driven array processing and array design techniques. Finally, to cope with irregular  virtual array geometries and wide-angle sensing requirements, we devise a universal and computationally efficient DOA estimation algorithm that remains applicable to arbitrary array configurations while significantly reducing computational cost.
	The main contributions of this work are summarized as follows:
	\begin{itemize}
		\item \textbf{\textit{Innovative Integrated Architecture:}} A novel FA-HAD architecture is proposed that integrates the advantages of FA array and HAD architecture. In this framework, an initial array configuration composed of multiple fluid antennas is organized into dynamically adjustable groups, enabling real-time array reconfiguration through coordinated motion. The spatially sampled signals are routed to RF chains via dynamic antenna selection, phase adjustment, and signal synthesis mechanisms. The fluid antenna array maximizes the utilization of spatial DOFs in constrained areas through dynamic spatial configuration, while the HAD processing mechanism significantly reduces pilot overhead and mitigates the complexity of high-dimensional signal processing through analog-domain signal compression. To the best of our knowledge, this is the first innovative architecture that integrates the hybrid analog-digital signal processing paradigm into a movable antenna system, providing a new theoretical framework to address the trade-off between dynamic array scalability and hardware complexity.
		
		\item \textbf{\textit{Fast DOA Estimation Strategy:}} Leveraging the proposed FA-HAD architecture, a proposed methodology employs single-point random phase shifting for omnidirectional spatial signal acquisition and achieves rapid DOA estimation through data compression characteristics, thereby markedly improving estimation timeliness and enhancing real-time system responsiveness.  Meanwhile, the achievable performance limits under compressive observations are characterized via single-source CRLB derivations, providing theoretical justification and quantitative support for the effectiveness of the proposed integrated architecture.
		
		\item \textbf{\textit{Efficient Virtual Array SCM Reconstruction:}} An efficient virtual-array SCM reconstruction method is developed to accurately recover a physically meaningful covariance representation for large-scale virtual arrays, thereby significantly enhancing multi-source resolvability. More importantly, the reconstructed SCM provides a covariance-domain interface that is compatible with a broad class of existing covariance-driven array processing and array design techniques, enabling seamless integration with well-established frameworks.
		
    	\item \textbf{\textit{Universal Low Complexity DOA Estimation:}} a novel dimension-reduced MUSIC algorithm based on the Jacobi-Anger polynomial expansion demonstrates exceptional array configuration adaptability, particularly well-suited for FAS implementations. Our solution attains high-precision DOA estimation while dramatically reducing computational overhead, offering an efficient and versatile approach for real-time angle estimation in complex operational environments.
	\end{itemize}

	\emph{Notations}: $\mathbf{A}$, $\mathbf{a}$ and $a$ represent matrix, vector and scalar, respectively. The superscripts $(\cdot)^{T}$ , $(\cdot)^{H}$, $(\cdot)^{-1}$ and $(\cdot)^{\dagger}$ denote the transpose, conjugate transpose, inverse and pseudo-inverse operators, respectively. $\mathbb{C}^{M\times N}$ and $\mathbb{R}^{M\times N}$ respectively are the sets of $\mathit{M\times N}$ complex-valued and real-valued matrices, while $\left\lbrace 0,1\right\rbrace^{M\times N}$ consists of $\mathit{M\times N}$-dimensional 0- and 1-valued matrices. $\odot$ stands for the Hadamard product, $j = \sqrt { - 1}$ the imaginary unit and $\mathbf{I}_N$ the $N\times N$ identity matrix. $\mathbb{E}\{  \cdot \} $ represents the statistical expectation, ${\mathop{\rm diag}\nolimits} \left\{  \cdot  \right\}$ the diagonalization operator, while $\|\cdot\|_F$, $\|\cdot\|_2$ and $\|\cdot\|_1$ the Frobenius norm, $\ell_2$ norm and $\ell_1$ norm, respectively. Finally $\alpha\sim\mathcal{U}(\rho, \upsilon)$ is a random variable uniformly distributed on the interval $\left[\rho, \upsilon\right]$, and ${\bf{n}}\sim\mathcal{CN}\left( {{\boldsymbol \mu},{\boldsymbol \Sigma}} \right)$ implies that ${\bf{n}}$ follows a complex Gaussian distribution with mean ${\boldsymbol \mu}$ and covariance ${\boldsymbol \Sigma}$. $\mathrm{Var}(\cdot)$ and $\mathrm{Cov}(\cdot,\cdot)$ denote the variance and covariance respectively.
	\section{System Overview}
	This section presents a comprehensive FA array-assisted HAD architecture specifically designed for DOA estimation. The proposed system strategically employs coordinated FAS movement in conjunction with HAD signal processing to achieve efficient spatial sampling while simultaneously reducing hardware implementation complexity.  
	
	\begin{figure*}[!t]
		\centering
		\includegraphics[width=2.0\columnwidth]{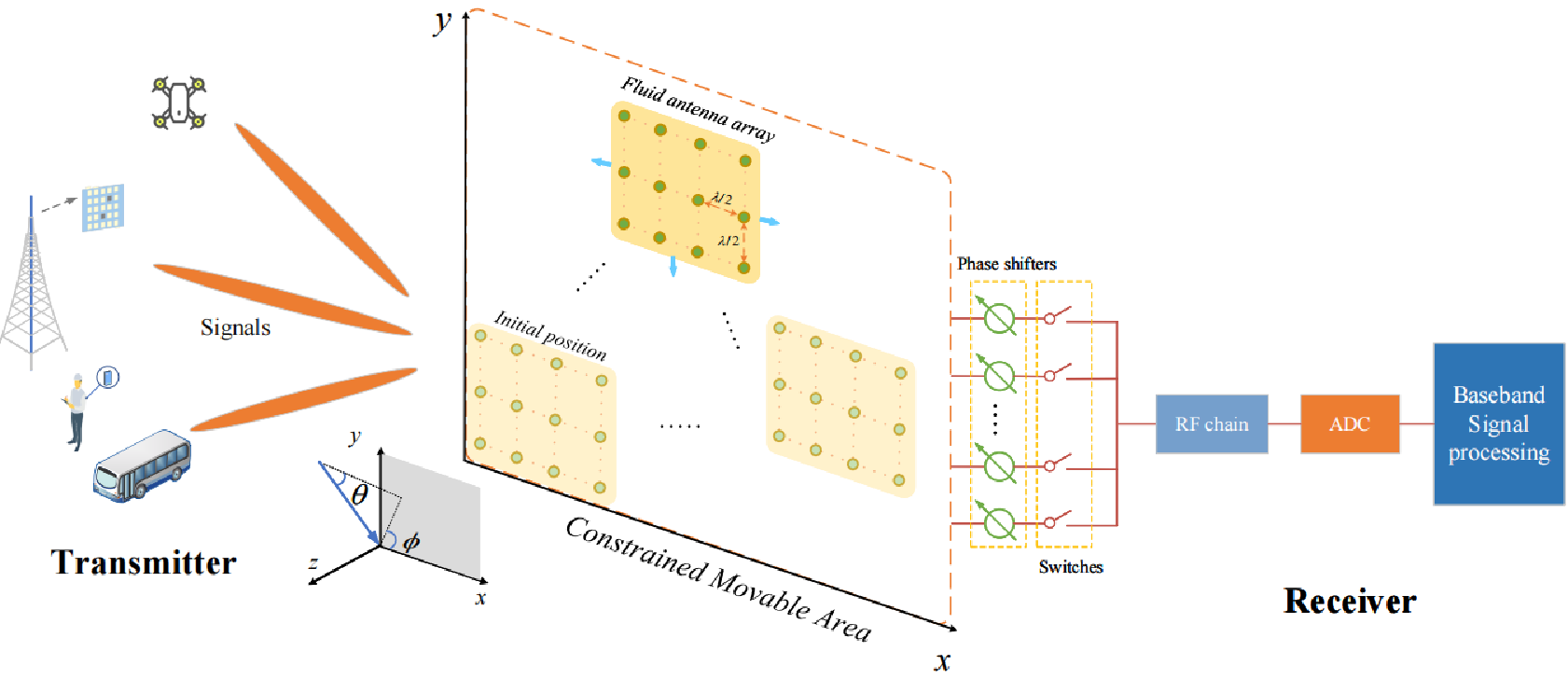}
		\caption{FA array-assisted HAD architecture.}
		\label{Fig.1}
	\end{figure*}
	
	\subsection{FA-HAD Architecture}
	Consider an intelligent wireless sensing architecture assisted by  FA array, as illustrated in Fig. \ref{Fig.1}. The receiver is equipped with an $N$-element FA array, with the ability for dynamic topology reconfiguration within a limited spatial domain, thereby providing multi-dimensional data for subsequent DOA estimation. 
	
	Unlike existing paradigms where each FA moves independently, this architecture  adopts a coordinated movement strategy that treats the entire array as an equivalent FA entity. \textit{This approach significantly enhances spatial sampling efficiency through synchronized displacement of intra-group elements.} In practice, such coordinated spatial reconfiguration can be realized through diverse physical mechanisms \cite{r43,r26}---including mechanically repositionable platforms, electronically switchable pixel arrays, or reconfigurable metasurfaces---without being restricted to any specific fluidic implementation. The essential requirement is that the effective radiating positions of all array elements can be jointly and synchronously adjusted within the designated spatial aperture. This coordinated movement mechanism eliminates the need for additional distributed control units, thereby effectively reducing overall system complexity.
	
	Furthermore, the receiver adopts an HAD architecture that dynamically compensates phase, selects signal paths, and aggregates energy for multiple received signals through coordinated control of phase shifters and switches. This processing ultimately outputs dimension-reduced composite signals to the RF chain for digital processing. Our design significantly reduces observation dimensionality while fully preserving both the aperture expansion advantages and high-resolution characteristics of the virtual array.

	%We consider the intelligent wireless sensing architecture assisted by a FA array as illustrated in Fig. \ref{Fig.1}. The base station (BS) is equipped with an FAS composed of $N$ FA, which achieves efficient spatial information acquisition through dynamic topology reconfiguration within a limited spatial domain, thereby providing multidimensional observation data for subsequent DOA estimation. Unlike existing paradigms where each FA moves independently, this architecture innovatively adopts a coordinated movement strategy, treating the entire array as an equivalent FA entity. This approach significantly enhances spatial sampling efficiency through synchronized displacement of intra-group elements. The hardware implementation of the array, as shown in Fig. \ref{Fig.1}, employs a one-dimensional tubular container integrated design, where conductive fluid is position-controlled via high-precision electronically regulated nanopumps. This coordinated movement mechanism eliminates the need for additional distributed control units, effectively reducing system complexity.  What's more, the receiver (Rx) adopts a HAD architecture, which dynamically compensates phase, selects paths, and aggregates energy for multiple received signals through coordinated control of phase shifters and switches. This ultimately outputs dimensionally compressed composite signals to the RF chain for digital processing. This design significantly reduces observation dimensionality while fully preserving the aperture expansion advantages and high-resolution characteristics of the virtual array.
	
	Assuming the system completes $K$ coordinated displacements, the received signal at the $k$-th spatial sampling position can be modeled as \footnote{Although modern hardware technologies are capable of enabling high-precision implementations, incorporating motion errors and front-end nonidealities, such as finite-resolution phase shifters and nonzero switching delay, into a more comprehensive system model constitutes an interesting direction for future research.}
	\begin{equation}\label{1}
		\mathbf{y}_k = \mathbf{w}^{H}_k\mathbf{H}_k\mathbf{S}+\mathbf{w}_k^{H}\mathbf{N}_k,
	\end{equation}
	where $\mathbf{S}\in\mathbb{C}^{{L\times N_p}}$ denotes the orthogonal pilot signal matrix transmitted by $L$ sensing targets, satisfying $\mathbf{S} \mathbf{S}^H = \mathbf{I}_L$. The additive noise matrix $\mathbf{N}_k\in\mathbb{C}^{{N\times N_p}}$ is assumed to be white Gaussian distributed, where each row follows $\mathcal{CN}(0, \sigma_n^{2} \mathbf{I}_{N})$. 
	
	The combining vector $\mathbf{w}_k\in\mathbb{C}^{{N\times 1}}$ is governed by the coordinated interplay between a switching network and an analog phase-shifting network, with its dynamic characteristics implicitly parameterized by the spatial sampling position index of the FA array. The unified mathematical representation can be modeled as
	\begin{equation}\label{2}
		\mathbf{w}_k = \mathbf{m}_{{s,k}} \odot\mathbf{m}_{{p,k}},
	\end{equation}
	where $\mathbf{m}_{{s,k}}\in \{0,1\}^{{N\times 1}}$ denotes the binary switching vector with each element being either 1 or 0. The phase-shifting vector $\mathbf{m}_{{p,k}} = \frac{1}{\sqrt{N}} \left[ e^{j\alpha_1}, e^{j\alpha_2}, \cdots, e^{j\alpha_{N}} \right]^T$ satisfies the constant-modulus constraint, where $0 \leq \alpha_{n} < 2\pi $ denotes the phase shift coefficient.

	%Assuming the system completes $K$ coordinated displacements, the received signal at the $k$-th spatial sampling positioncan be modeled as
	%	\begin{equation}\label{1}
		%	\mathbf{y}_k = \mathbf{w}^{H}_k\mathbf{H}_k\mathbf{S}+\mathbf{w}_k^{H}\mathbf{N}_k
		%	\end{equation}
	%	where $\mathbf{S}\in\mathbb{C}^{{L\times N_p}}$ denotes the orthogonal pilot signals transmitted by $L$ sensing targets, satisfying $\mathbf{S} \mathbf{S}^H = \mathbf{I}_L$.  Furthermore, the additive noise $\mathbf{N}_k\in\mathbb{C}^{{N\times N_p}}$ is assumed to be Gaussian white one, whose each row obeys $\mathbf{n}\sim\mathcal{CN}(0, \sigma_n^{2} \mathbf{I}_{N})$. The analog combining vector $\mathbf{w}_k\in\mathbb{C}^{{N\times 1}}$ is governed by the coordinated interplay of a switching network and a phase-shifting network, with its dynamic characteristics implicitly parameterized by the spatial sampling position index of the FA array. The unified mathematical representation can be formulated as
	%	\begin{equation}\label{2}
		%		\mathbf{w} = \mathbf{m}_{\mathrm{s}} \odot\mathbf{m}_{\mathrm{p}}
		%	\end{equation}
	%	where $\mathbf{m}_{\mathrm{s}}\in \{0,1\}^{{N\times 1}}$ denotes the binary switch with each element being either 1 or 0. The phase-shifting vector $\mathbf{m}_{\mathrm{p}} = \frac{1}{\sqrt{N}} \left[ e^{j\alpha_1}, e^{j\alpha_2}, \cdots, e^{j\alpha_{N}} \right]^T$ satisfies the constant-modulus constraint and  $0 \leq \alpha_{n} \leq 2\pi $ denotes the phase shift coefficient.

	\subsection{Channel Model}
	To ensure model generality within a rigorous theoretical framework, this study establishes the channel model based on the following fundamental assumptions: 
	\begin{enumerate}
		\item Line-of-sight (LOS) propagation dominates the transmission scenario;
		\item The far-field approximation is valid, where the FA array's displacement range is significantly smaller than the electromagnetic wave propagation distance;
		\item Quasi-static fading characteristics are considered, implying that both the DOAs and complex gain coefficients of all propagation paths remain invariant during the dynamic reconfiguration of the FA array.
	\end{enumerate}
	Under these assumptions, the transmitter-receiver channel matrix at the $k$-th spatial sampling position can be expressed as:
	%Within the theoretical framework ensuring model generality, this study establishes the channel model under the following fundamental assumptions: (1) Line-of-sight (LOS) propagation dominates the transmission scenario; (2) The far-field approximation holds, where the FA array's displacement range is significantly smaller than the electromagnetic wave propagation distance; (3) Quasi-static fading characteristics are assumed, implying that the DOAs and complex gain coefficients of all propagation paths remain invariant during the dynamic reconfiguration of the fluid antenna array. Under these assumptions, the transmitter-receiver channel matrix at the $k$-th spatial sampling position can be expressed as:
	\begin{equation}\label{3}
		\mathbf{H}_k = \mathbf{A}_{k}\mathbf{\Lambda},
	\end{equation}
	where $\mathbf{\Lambda}= \mathrm{diag}\{\gamma_1,\cdots,\gamma_L\}$ is the diagonal matrix consisting of the complex path gains of the multipath channel. $\mathbf{A}_{k}$ denotes the dynamic array manifold matrix of the FA array at the $k$-th spatial sampling position, which can be expressed as
	%where $\mathbf{\Lambda}= \mathrm{diag}\{\gamma_1,\cdots,\gamma_L\}$ consists of the path gains of complex multipath channel. $\mathbf{A}_{k}$ represents the dynamic array manifold matrix of the FA array at the $k$-th spatial sampling position, which can be written as
	\begin{equation}\label{4}
		\mathbf{A}_k=\left[\mathbf{a}_k(\theta_1, \phi_1),\cdots,\mathbf{a}_k(\theta_L, \phi_L)\right],
	\end{equation}
	where $\theta_\ell$ and $\phi_\ell$ represent the azimuth and elevation angles of the $\ell$-th path, respectively. $\mathbf{a}_k(\theta_\ell, \phi_\ell)$ characterizes the steering vector of the FA array, whose spatial response characteristics are jointly determined by both the initial topological configuration and the dynamic displacement positions of the array.
	
	%where $\theta_\ell$ and $\phi_\ell$ are the azimuth and elevation angle of the $\ell$-path. Vector $\mathbf{a}_k(\theta_\ell, \phi_\ell)$ characterizes the steering vector of the FA array, whose spatial response characteristics are jointly determined by the initial topological configuration and dynamic displacement positions of the array. 
	%The construction of steering vectors fundamentally depends on the physical distance relationships between array elements and the reference element. The physically reconfigurable characteristics of FAs inherently break the geometric constraints of conventional arrays. Let $(x_n,y_n,0)$ represent the initial spatial position of the $n$-th antenna in the three-dimensional Cartesian coordinate system.  The steering vector of FA array in its initial position  can be expressed as 
	
	The construction of steering vectors fundamentally depends on the physical distance relationships between array elements and a reference element. The physically reconfigurable characteristics of FAS inherently transcend the geometric constraints of conventional arrays. Let $(x_n,y_n,0)$ represent the initial spatial coordinates of the $n$-th antenna in the 3-D Cartesian coordinate system. The steering vector of the FA array at its initial position can be expressed as
	\begin{multline}\label{5}
		{\mathbf{a}_1(\theta, \phi)=}[1,\cdots,e^{j2\pi (x_n\sin\theta\cos\phi+y_n\sin\theta\sin\phi)/\lambda}, \\
		\cdots, e^{j2\pi (x_N\sin\theta\cos\phi+y_N\sin\theta\sin\phi)/\lambda}{]^{T}}.
	\end{multline}
	
	By leveraging the array manifold matrix $\mathbf{A}_1$ of the FA array in its initial configuration, $\mathbf{A}_k$ can be mathematically written as
	\begin{equation}\label{6}
		\mathbf{A}_k =\mathbf{A}_1\boldsymbol{\Psi}_k
	\end{equation}
	where $\boldsymbol{\Psi}_k\in\mathbb{C}^{{L\times L}}$ characterizes the phase difference between the received signals at the initial position and the $k$-th displaced position, given by
	\begin{equation}\label{7}
		\boldsymbol{\Psi}_k = \mathrm{diag} \left\{ e^{j \frac{2 \pi}{\lambda} (d_{k, x} \beta_{1,x} + d_{k,y} \beta_{1,y})}, \ldots, e^{j \frac{2 \pi}{\lambda} (d_{k, x} \beta_{L,x} + d_{k,y} \beta_{L,y})} \right\}
	\end{equation}
	where $d_{k,x}$ and $d_{k,y}$ denote the geometric displacement components relative to the initial position along the $x$- and $y$-axes, respectively; $\beta_{\ell,x} = \sin\theta_\ell\cos\phi_\ell$ and	$\beta_{\ell,y} = \sin\theta_\ell\cos\phi_\ell$ are the direction-cosine components of the $\ell$-th impinging signal direction along the $x$- and $y$-axes, respectively.

	%	By leveraging the array manifold matrix $\mathbf{A}_1$ of the FA array in its initial configuration, $\mathbf{A}_k$ can be mathematically formulated as
	%	\begin{equation}\label{6}
		%		\mathbf{A}_k =\mathbf{A}_1\boldsymbol{\Psi}_k
		%	\end{equation}
	%	where $\boldsymbol{\Psi}_k\in\mathbb{C}^{{L\times L}}$ characterizes the phase difference between the received signals at the initial position and the $k$-th displaced position. Given that the FA array is constrained to unidirectional displacement along the $x$-axis in the Cartesian coordinate system, $\boldsymbol{\Psi}_k$ can be written as
	%	\begin{equation}\label{7}
		%		\boldsymbol{\Psi}_k = \mathrm{diag} \left\{ e^{j 2 \pi \frac{d_k}{\lambda} \beta_1}, \ldots, e^{j 2 \pi \frac{d_k}{\lambda} \beta_L} \right\}
		%	\end{equation}
	%	where $\beta_\ell = \sin\theta_\ell\cos\phi_\ell$ and and $d_k$ is the distance from the initial position.	
	\section{Collaborative Sampling and Performance Bounds}
	This section proposes a coordinated spatial-phase sampling mechanism tailored to the FA-HAD architecture. By introducing complementary observation diversity along both spatial reconfiguration and phase reconfiguration, the proposed strategy preserves essential angular information and enables fast 2-D DOA estimation under compressive observations with a limited number of RF chains. Meanwhile, the achievable performance of this strategy is theoretically characterized, providing quantitative guidance for system-parameter design and the accuracy-overhead trade-off.
	
	%This section presents a 2D direction-of-arrival estimation framework for FAS, addressing key spatial sensing challenges through three innovations: a coordinated spatial phase sampling mechanism for efficient omnidirectional channel acquisition, a virtual array covariance reconstruction algorithm to mitigate compressed observation losses, and a Jacobi-Anger expansion-based reduced-dimension MUSIC algorithm for high-precision angle estimation with irregular arrays.  Systematic complexity analysis confirms the proposed methods enhances computational efficiency without sacrificing accuracy, supporting practical reconfigurable antenna system implementations.
	\subsection{Collaborative Spatial-Phase Acquisition and Data Fusion}
	
	Compared with 1-D DOA estimation using a uniform linear array (ULA), 2-D DOA estimation requires resolving both azimuth and elevation, where the equivalent phase-observation space carries richer angular information and thus demands higher observation diversity. To preserve essential angular features under compressive observations with a limited number of RF chains, a collaborative spatial-phase  sampling mechanism is developed. Specifically, the spatial reconfiguration of the FA array introduces geometric variations that expand the available spatial DOFs, thereby yielding richer observation information; meanwhile, at each fixed spatial location after movement, multiple analog combining vectors are employed to generate distinct compressed projections of the received signals, thereby providing more discriminative observation representations. The collaboration of these two dimensions enables omnidirectional angular excitation without any prior DOA information, thereby establishing an informative yet implementation-friendly basis for subsequent compressive-domain DOA estimation.
	
	Specifically, at the $k$-th spatial sampling position, $T$ pilot observations are collected using $T$ combining vectors $\mathbf{w}_k^{\tau}$, where the phase shift coefficients $\{\alpha_n\}$ in \eqref{2} are generated from  $\mathcal{U}[0, 2\pi)$. This randomized phase design is not intended to be mathematically optimal; rather, it serves a prior-free generalized beam-alignment principle. By applying multiple independent phase modulations, a non-directional excitation over the angular domain is created, yielding sufficient phase diversity and broad angular coverage with relatively low hardware and pilot overhead. As a result, the proposed design is general and practical, insensitive to the specific mobility pattern and phase-shifter accuracy, and allows spatial and phase sampling to work collaboratively to support fast 2-D DOA estimation under compressive measurements.

	%Compared with one-dimensional DOA estimation using uniform linear arrays (ULA), 2-D DOA estimation inherently contains richer spatial angular information within the same phase observation space, necessitating higher phase resolution precision. To comprehensively capture omnidirectional channel components, the proposed FA array employs $T$ randomly generated combining vectors at each spatial sampling position. Specifically, the $\tau$-th training combining vector at the $k$-th array position is denoted as $\mathbf{w}^\tau_k$, whose phase shift coefficients $\alpha_n$ in \eqref{2} are randomly generated from a uniform distribution $\alpha_n\sim\mathcal{U}[0, 2\pi)$.
	
	To establish a tractable mathematical framework, we define the equivalent transmitted signal matrix as $\mathbf{\bar S}=\mathbf{\Lambda} \mathbf{S}$ \footnote{This study primarily focuses on developing effective DOA estimation algorithms while deliberately deferring the channel gain acquisition problem. The theoretical justification stems from the fact that channel gain coefficients do not influence the statistical characteristics of sensing target signals, thereby making it mathematically sound to embed these coefficients into the equivalent transmitted signal model.}. Based on this formulation, the observed signal corresponding to the combining vector $\mathbf{w}^\tau_k$ can be expressed according to \eqref{1} as
	\begin{equation}\label{8}
		\mathbf{y}^{\tau}_k = {\mathbf{w}^{\tau}_k}^H\mathbf{A}_k\mathbf{\bar S}+{\mathbf{w}^{\tau}_k}^H\mathbf{N}^{\tau}_k.
	\end{equation}
	
	%Through the definition of the equivalent transmitted signal as $\mathbf{\bar S}=\mathbf{\Lambda} \mathbf{S}$ \footnote{This study primarily focuses on developing an effective DOA estimation algorithm, while intentionally deferring the channel gain acquisition problem. The theoretical justification lies in the fact that channel gain coefficients do not affect the statistical characteristics of the sensing target signals, making it theoretically justified to embed these coefficients into the equivalent transmitted signal model.}, the observed signal corresponding to the combining vector $\mathbf{w}^\tau_k$ can be formulated according to \eqref{1} as
	%\begin{equation}\label{8}
	%	\mathbf{y}^{\tau}_k = {\mathbf{w}^{\tau}_k}^H\mathbf{A}_k\mathbf{\bar S}+{\mathbf{w}^{\tau}_k}^H\mathbf{N}^{\tau}_k.
	%\end{equation}
	
	Aggregating the observed signals obtained from $K$ spatial sampling locations in the $\tau$-th omnidirectional channel components acquisition, we obtain
	\begin{equation}\label{9}
		\mathbf{Y}_\tau = \mathbf{W}_\tau^{H}\mathbf{A}\mathbf{\bar S}+\mathbf{W}_\tau^{H}\mathbf{N}_\tau,
	\end{equation}
	where $\mathbf{Y}_\tau = \left[ {\mathbf{y}^\tau_1}^T,\ldots, {\mathbf{y}^\tau_K}^T \right]^T\in \mathbb{C}^{{K\times N_p}}$, $\mathbf{N}_\tau = \left[ {\mathbf{N}^\tau_1}^T,\ldots, {\mathbf{N}^\tau_K}^T \right]^T\in \mathbb{C}^{{NK\times N_p}}$, $\mathbf{A}= \left[ \mathbf{A}_1^T,\ldots, \mathbf{A}_K^T \right]^T$ and $\mathbf{W}_\tau = \mathrm{diag} \left\{\mathbf{w}^{\tau}_1,\ldots,\mathbf{w}^{\tau}_K\right\}\in \mathbb{C}^{{NK\times K}}$, respectively.
	The array manifold matrix $\mathbf{A}\in \mathbb{C}^{{NK\times L}}$
	of the virtual array comprehensively embodies the spatial-polarimetric characteristics of the FA system. It is explicitly parameterized as
	\begin{equation}\label{10}
		\mathbf{A}=\left[\mathbf{a}(\theta_1, \phi_1),\cdots,\mathbf{a}(\theta_L, \phi_L)\right],
	\end{equation}
	where $\mathbf{a}(\theta_\ell, \phi_\ell) = \left[ \mathbf{a}_1(\theta_\ell, \phi_\ell)^T,\ldots, \mathbf{a}_K(\theta_\ell, \phi_\ell)^T \right]^T\in\mathbb{C}^{{KN\times 1}}$. 
	By defining $\mathbf{\bar N}_\tau=\mathbf{W}_\tau^{H}\mathbf{N}_\tau$ as modified noise matrix, the $T$-time sequential spatial scanning can be compiled into a unified spatiotemporal data set through multi-dimensional signal stacking as
	\begin{equation}\label{11}
		\mathbf{Y}
		=\mathbf{W}^{H}\mathbf{A}\mathbf{\bar S}+\mathbf{\bar N},
	\end{equation}
	where $\mathbf{Y} = \left[ \mathbf{Y}_1^T,\ldots, \mathbf{Y}_T^T \right]^T\in \mathbb{C}^{{KT\times N_p}}$, $\mathbf{\bar N} = \left[ \mathbf{\bar N}_1^T,\ldots, \mathbf{\bar N}_T^T \right]^T\in \mathbb{C}^{{KT\times N_p}}$ and $\mathbf{W} = \left[ \mathbf{W}_1^T,\ldots, \mathbf{W}_T^T \right]^T\in \mathbb{C}^{{NK\times KT}}$.
	\begin{table}[!t]
		\centering 
		\caption{Comparison of Different Antenna System Architectures}
		\label{tab:architecture_comparison}
		\begin{tabular}{|l|c|c|c|c|}
			\hline
			\textbf{Architecture} & \textbf{Antenna} & \textbf{RF Chain} & \textbf{Mechanical} & \textbf{Pilot} \\
			\textbf{Paradigm} & \textbf{Number} & \textbf{Number} & \textbf{Adjustments} & \textbf{Overhead} \\
			\hline
			Single FA & 1 & 1 & $NK$ & $NK$ \\
			\hline
			Fully Digital & $N$ & $N$ & $K$ & $K$ \\
			FA Array & & & & \\
			\hline
			\textbf{Proposed} & $N$ & $1$ & $K$ & $TK$ \\
			\textbf{FA-HAD} & & & & $T < N$ \\
			\hline
		\end{tabular}
	\end{table}
	\subsection{FA-HAD-MUSIC Algorithm}
	Given $\mathbf{Y}\in\mathbb{C}^{{KT\times N_p}}$, the SCM of the phase-compressed synthetic observations is calculated as
	\begin{equation}\label{12}
		\mathbf{R}_\mathrm{com} = \frac{1}{N_p} \mathbf{Y} \mathbf{Y}^H  =\mathbf{E} \mathbf{R}_s\mathbf{E}^H + \mathbf{R}_n,
	\end{equation}
	where $\mathbf{R}_s=\frac{1}{N_p} \mathbf{\bar S} \mathbf{\bar S}^H$ and $\mathbf{R}_n = \frac{1}{N_p} \mathbf{\bar N} \mathbf{\bar N}^H$ represent the covariance matrices of the equivalent signal $\mathbf{\bar S}$ and modified noise $\mathbf{\bar N}$, respectively, and {$\mathbf{E} =\mathbf{W}^H\mathbf{A}$} denotes the augmented array manifold matrix.
	
	Performing eigenvalue decomposition (EVD) on $\mathbf{R}_\mathrm{com}$ yields
	\begin{equation}\label{13}
		\mathbf{R}_\mathrm{com} =\mathbf{U}_s \mathbf{\Sigma}_s \mathbf{U}_s^{H} + \mathbf{U}_n \mathbf{\Sigma}_n \mathbf{U}_n^{H},
	\end{equation}
	where the first and second components correspond to the signal and noise subspaces, respectively. Here,  $\mathbf{\Sigma}_s\in \mathbb{C}^ {L\times L}$ and $\mathbf{\Sigma}_n\in \mathbb{C}^{(KT-L)\times (KT-L)}$ are two diagonal matrices that consist of $L$ largest eigenvalues and remaining eigenvalues, respectively. $\mathbf{U}_s\in \mathbb{C}^{KT\times L}$ and $\mathbf{U}_n \in \mathbb{C}^{KT \times (KT - L)}$ contain eigenvectors corresponding to $\mathbf{\Sigma}_s$ and $\mathbf{\Sigma}_n$, respectively.\footnote{{The number of signal sources $L$ is assumed to be known in this work, which can be obtained via prior information or standard source enumeration algorithms, such as the Akaike information criterion (AIC) \cite{r38}, minimum description length (MDL) criterion \cite{r39}, and second-order difference (SOD) operation on least-squares coefficients \cite{r40}.}}
	
	Leveraging the inherent orthogonality between array manifold vectors and the noise subspace, the spectrum estimation function is
	\begin{equation}\label{14}
		P(\theta,\phi)=\frac{1}{{\bf{a}}^H(\theta ,\phi ){\bf E}_{n}{\bf E}_{n}^H{\bf{a}}({\bf{\theta }},\phi)},
	\end{equation}
	where ${\bf E}_{n}={{\bf W}}{{\bf U}_{n}}$ represents the augmented noise subspace. The DOA estimates $(\hat \theta_\ell ,\hat \phi_\ell)$ for sensing targets can be determined by identifying $L$ prominent local maximum of the normalized pseudospectrum function through a 2-D angular search. The FA-HAD-MUSIC for 2-D DOA estimation is summarized in \textbf{Algorithm \ref{al1}}.

	\emph{Remark 1:} Other existing studies on FAS adopt a single FA design paradigm, which necessitates $NK$ mechanical adjustments and equivalent pilot resources to sample $NK$ spatial position states. In stark contrast, the proposed spatiotemporal joint sampling framework achieves substantial efficiency improvements by requiring only $K$ array adjustments while reducing pilot overhead to $TK$ (where $T < N$). This methodological advancement effectively alleviates the inherent trade-off between hardware response rate and spectral efficiency that has been a persistent challenge in large-scale array system implementations.
	
	Table \ref{tab:architecture_comparison} provides a comparison of the proposed FA-HAD architecture against existing approaches, clearly demonstrating the significant efficiency gains achieved through the HAD paradigm.
\begin{algorithm}[!t]
	\caption{FA-HAD-MUSIC for FAS-aided 2-D DOA Estimation}
	\label{al1}
	\LinesNumbered
	\KwIn{Multi-frame observations $\{\mathbf{y}^\tau_k\}_{k=1,\tau=1}^{K,T}$, combining vectors $\{\mathbf{w}^\tau_k\}_{k=1,\tau=1}^{K,T}$}
	\KwOut{DOA estimates $\{(\hat\theta_\ell,\hat\phi_\ell)\}_{\ell=1}^L$}
	
	\For{$\tau=1$ \KwTo $T$}{
		\textbf{Spatial-location fusion:} 
		$\mathbf{Y}_\tau \leftarrow \mathrm{stack}(\mathbf{y}^\tau_1,\ldots,\mathbf{y}^\tau_K)$,
		$\mathbf{W}_\tau = \mathrm{diag}\{\mathbf{w}^\tau_1,\ldots,\mathbf{w}^\tau_K\}$,
	}
	\textbf{Spatial-phase fusion:}
	$\mathbf{Y} \leftarrow \mathrm{stack}(\mathbf{Y}_1,\ldots,\mathbf{Y}_T)$,
	 $\mathbf{W} \leftarrow \mathrm{stack}(\mathbf{W}_1,...,\mathbf{W}_T)$,
	
	Compute fused covariance:
	$\mathbf{R}_{\mathrm{com}} = \frac{1}{N_p}\mathbf{Y}\mathbf{Y}^H$,
	
	EVD and augmented subspace separation:
	$\mathbf{R}_{\mathrm{com}} \xrightarrow{\mathrm{EVD}} \{\mathbf{U}_s,\mathbf{U}_n\}$,
	
	Precompute noise projector:
	$\mathbf{W}\mathbf{U}_n\mathbf{U}_n^H\mathbf{W}^H$,
	
	\textbf{Classical 2-D MUSIC:}
	Construct $\mathbf{a}(\theta,\phi)$ and evaluate
	\eqref{14} over the search grid; 
	select the $L$ largest peaks to obtain $\{(\hat\theta_\ell,\hat\phi_\ell)\}_{\ell=1}^L$,
	
	\Return $\{(\hat\theta_\ell,\hat\phi_\ell)\}_{\ell=1}^L$.
\end{algorithm}

	\subsection{Cram\'{e}r-Rao Lower Bound}
    The CRLB provides a fundamental lower bound on the covariance of any unbiased estimator. To facilitate the subsequent derivation, all deterministic but unknown parameters are collected into the parameter vector as
	\begin{equation}
			\boldsymbol{\eta} = \left[\boldsymbol{\theta}^T,\ \mathbf{\bar{s}}^T(1),\ldots, \mathbf{\bar{s}}^T(N_p),\sigma_n^2\right]^T,
	\end{equation}
	where $\boldsymbol{\theta} = \left[\theta_1,\ldots,\theta_L,\phi_1,\ldots,\phi_L\right]^T$ contains the set of two-dimensional angular parameters to be estimated. Accordingly, the covariance matrix of any unbiased estimator $\hat{\boldsymbol{\eta}}$ satisfies
		\begin{equation}
			\mathbb{E}\!\left[
			\left(\hat{\boldsymbol{\eta}}-\boldsymbol{\eta}\right)
			\left(\hat{\boldsymbol{\eta}}-\boldsymbol{\eta}\right)^H
			\right]\geq
			{\mathop{\rm CRLB}\nolimits},
		\end{equation}
		
		Based on the aggregated compressed observation data obtained from \eqref{11}, i.e., $\mathbf{Y} = \left[\mathbf{y}(1),\ldots,\mathbf{y}(N_p)\right]$, the observation signal can be regarded as following a complex Gaussian distribution with mean vector $\boldsymbol{\mu}$ and covariance matrix $\boldsymbol{\Gamma}$, given by
		\begin{equation}
			\boldsymbol{\mu} =
			\begin{bmatrix}
				\pmb{\Phi}\mathbf{A}\mathbf{\bar s}(1) \\
				\vdots \\
				\pmb{\Phi}\mathbf{A}\mathbf{\bar s}(N_p)
			\end{bmatrix}\quad
			\boldsymbol{\Gamma} =
			\begin{bmatrix}
				\sigma_n^2 \mathbf{Q} & \cdots & \mathbf{0} \\
				\vdots & \ddots & \vdots \\
				\mathbf{0} & \cdots & \sigma_n^2 \mathbf{Q}
			\end{bmatrix},
		\end{equation}
		where $\pmb{\Phi}$ denotes the augmented analog combining matrix $\mathbf{W}^H$. For notational convenience, we further define its correlation matrix as $\mathbf{Q}=\pmb{\Phi}\pmb{\Phi}^H \in \mathbb{C}^{KT \times KT}$. On this basis, the Fisher information matrix (FIM) can be established from the log-likelihood function associated with the complex Gaussian model. Consequently, the $(i,j)$-th entry of the FIM is given by
		\begin{equation}
			[\pmb{\mathcal{F}}]_{i,j} = 2\mathcal{R}\left\{ \frac{\partial \boldsymbol{\mu}^H}{\partial \eta_i} \boldsymbol{\Gamma}^{-1} \frac{\partial \boldsymbol{\mu}}{\partial \eta_j} \right\} + \text{tr} \left( \boldsymbol{\Gamma}^{-1} \frac{\partial \boldsymbol{\Gamma}}{\partial \eta_i} \boldsymbol{\Gamma}^{-1} \frac{\partial \boldsymbol{\Gamma}}{\partial \eta_j} \right),
		\end{equation}
	
	Once the FIM is derived, the CRLB can be readily obtained from the inverse of the corresponding information matrix. Building upon the CRLB analysis for 1-D DOA estimation in \cite{r41}, we generalize the underlying framework to the 2-D case under the proposed FA-HAD architecture. Consequently, the $2L \times 2L$ CRLB matrix for the 2-D DOA angle parameters is obtained as
	
	\begin{equation}\label{42}
		{\mathop{\rm CRLB}\nolimits} = \frac{\sigma_n^2}{2} \left[ \mathcal{R}\left\lbrace \sum\limits_{n = 1}^{{N_p}} \mathbf{\tilde{S}}^{H}(n)\mathbf{\tilde{B}}^{H} \pmb{\Pi}_{\mathbf{\tilde{A}}} \mathbf{\tilde{B}}\mathbf{\tilde{S}}(n)\right\rbrace \right]^{-1},
	\end{equation}
	where
	\begin{equation}\label{43}
		\mathbf{\tilde{B}} = \mathbf{Q}^{-\frac{1}{2}}\pmb{\Phi}\mathbf{B},
	\end{equation}
	\begin{equation}\label{44}
		\mathbf{B}= \left[ \frac{\partial \mathbf{a}}{\partial \theta} \bigg|_{\theta = \theta_1}, \dots, \frac{\partial \mathbf{a}}{\partial \theta} \bigg|_{\theta = \theta_L},\frac{\partial \mathbf{a}}{\partial \phi} \bigg|_{\phi = \phi_1},\dots, \frac{\partial \mathbf{a}}{\partial \phi} \bigg|_{\phi = \phi_L}\right],		
	\end{equation}
	\begin{equation}\label{45}
		\pmb{\Pi}_{\mathbf{\tilde{A}}} = \mathbf{I}_{K T} - \mathbf{\tilde{A}} \left( \mathbf{\tilde{A}}^H \mathbf{\tilde{A}} \right)^{-1} \mathbf{\tilde{A}}^H,
	\end{equation}
	\begin{equation}\label{46}
		\mathbf{\tilde{A}}=\mathbf{Q}^{-\frac{1}{2}}\pmb{\Phi}\mathbf{A},
	\end{equation}
	\begin{equation}\label{47}
		\mathbf{\tilde{S}}(n) = \mathrm{diag}\{\bar s_1(n),\cdots,\bar s_L(n),\bar s_1(n),\cdots,\bar s_L(n)\},
	\end{equation}
	with $\mathbf{a}=\mathbf{a}(\theta, \phi)\in\mathbb{C}^{KN\times 1}$ denoting the \emph{virtual array} steering vector, i.e., the generic column of the manifold matrix in \eqref{10},  $\mathbf{\tilde{S}}(n)\in\mathbb{C}^{ 2L\times 2L}$ the  diagonal matrix formed by the $n$-th column of   $\mathbf{\bar S}$, and $\bar s_\ell(n)$ the element in the $\ell$-th row and $n$-th column of   $\mathbf{\bar S}$.
	
	\subsection{Single Source Analysis}
     Although \eqref{42} provides the CRLB in a general form, its dependence on the system configuration is embedded in the coupled terms of the array manifold and the compressive combining operation, making it difficult to directly discern how key parameters govern information accumulation and the resulting error floor. To improve interpretability and highlight the dominant factors, we next specialize to the single-source case and explicitly expand the Fisher information matrix, yielding an analyzable approximation that more transparently reveals how the sampling and compression mechanisms shape the achievable performance bound.
       
	 The $2 \times 2$ Fisher information matrix $\pmb{\mathcal{F}}$ corresponding to \eqref{42} is rewritten as
	 \begin{equation}\label{76}
	 	\pmb{\mathcal{F}}
	 	={\mathop{\rm CRLB}\nolimits}^{-1} = \mathcal{R}\left\lbrace\begin{bmatrix}
	 		\mathbf{F}_{\pmb{\theta}\pmb{\theta}} & \mathbf{F}_{\pmb{\theta}\pmb{\phi}} \\
	 		\mathbf{F}_{\pmb{\phi}\pmb{\theta}} & \mathbf{F}_{\pmb{\phi}\pmb{\phi}}
	 	\end{bmatrix}\right\rbrace,
	 \end{equation}
	where
   	\begin{equation}\label{F_var_theta}
   	\mathbf{F}_{\pmb{\theta}\pmb{\theta}}
   	\approx \frac{8N_p \hat{p}\pi^2\cos^2\theta}{ \sigma_n^2\lambda^2}\, T K \,\mathrm{Var}\!\big(\bar d_{\dot n}\big).
   \end{equation}
   
   \begin{equation}\label{F_var_phi}
   	\mathbf{F}_{\pmb{\phi}\pmb{\phi}}
   	\approx \frac{8N_p \hat{p}\pi^2\sin^2\theta}{ \sigma_n^2\lambda^2}\, T K \,\mathrm{Var}\!\big(\tilde d_{\dot n}\big).
   \end{equation}
   
   \begin{equation}\label{F_cov}
   	\mathbf{F}_{\pmb{\theta}\pmb{\phi}}=\mathbf{F}_{\pmb{\phi}\pmb{\theta}}
   	\approx \frac{8N_p \hat{p}\pi^2\sin\theta\cos\theta}{ \sigma_n^2\lambda^2}\, T K \,\mathrm{Cov}\!\big(\bar d_{\dot n},\tilde d_{\dot n}\big).
   \end{equation}
    
    \begin{proof}
    	See Appendix A.
    \end{proof}
    
    	\emph{Remark 2:} From \eqref{F_var_theta}--\eqref{F_cov}, the parameter $T$ enters the Fisher information only through a linear scaling factor. This indicates that increasing the number of phase observations collected at each spatial location improves the phase diversity and correspondingly reduces the CRLB approximately in proportion to $1/T$. However, the role of $K$ is not merely to increase the number of measurements. More importantly, increasing the number of spatial samplings reshapes the virtual-element coordinate set and thereby modifies $\mathrm{Var}(\bar d_{\dot n})$, $\mathrm{Var}(\tilde d_{\dot n})$, and $\mathrm{Cov}(\bar d_{\dot n},\tilde d_{\dot n})$. Since these terms jointly characterize the aperture dispersion of the virtual array along the two angle-sensitive dimensions, enlarging $K$ typically yields a more pronounced improvement in estimation accuracy.
    	
    	\emph{Remark 3:} Under the adopted approximations, the contribution of $N$ does not directly dominate the Fisher information scale. Its gain is primarily geometry-dependent, namely, whether increasing $N$ can effectively enlarge or reshape the array aperture. If $N$ merely increases the element density within an essentially fixed aperture, the resulting theoretical improvement is generally limited and may gradually saturate. Even when the theoretical gain from increasing $N$ is limited, a larger initial array size still offers an important practical advantage under a prescribed spatial sampling trajectory. Specifically, it can acquire the required spatial DOFs with fewer movements and lower pilot overhead while maintaining the desired spatial coverage, thereby improving the DOFs-harvesting efficiency and deployment efficiency of the proposed system.
	
	\section{Covariance-Domain DOA Estimation}
	This section establishes a covariance-domain DOA estimation framework for the proposed FA-HAD architecture. By reconstructing a physically meaningful virtual-array SCM from compressive measurements, the framework restores a standard covariance representation that interfaces naturally with a broad class of covariance-based array processing and array design methods, thereby improving architecture scalability. 
	On this basis, a geometry-agnostic and computationally efficient 2-D DOA estimator is further developed using the Jacobi-Anger expansion to support reliable estimation for arbitrary planar arrays.
	\subsection{Virtual Array SCM Reconstruction}
	The fast FA-HAD-MUSIC strategy relies on reduced-dimensional compressive observations, which, while enabling rapid estimation, may limit the effective observation dimension and thus the number of resolvable sources, especially when the spatial sampling budget is small. More importantly, many well-established high-resolution DOA estimators and array-design methodologies are naturally formulated in the covariance domain, where a physically meaningful SCM serves as a fundamental statistical interface.
	
	Motivated by the above, we develop an efficient virtual-array SCM reconstruction method that recovers the covariance of the equivalent virtual array induced by FA movements, thereby restoring a physically meaningful covariance representation for subsequent high-resolution processing and facilitating compatibility with covariance-based array processing and design frameworks.
	
	%While FA-HAD-MUSIC algorithm significantly improves estimation efficiency through dimension-compressed observations, its random phase-shift mechanism introduces two critical limitations: degraded signal-to-noise ratio (SNR) causing DOA estimation bias under low-SNR conditions, and insufficient compressed signal dimensions (when the array movement count $K$ is small) restricting identifiable target quantity; to address these, we propose a high-precision sample covariance matrix recovery method based on virtual array reconstruction, which recovers the equivalent virtual array's covariance matrix generated by FA array movements, thereby enabling robust high-accuracy DOA estimation.

	According to \eqref{1} and \eqref{3}, the received signal of the FA array before analog combining at the $k$-th spatial sampling position can be expressed as
	\begin{equation}\label{15}
		\mathbf{\Upsilon}_k = \mathbf{H}_k\mathbf{S}+\mathbf{N}_k = \mathbf{A}_k\mathbf{\bar S}+\mathbf{N}_k.
	\end{equation}
	By concatenating the received signals across all $K$ spatial sampling positions, the equivalent large-scale virtual array signal with extended aperture can be constructed as $\mathbf{\Upsilon} = \left[ \mathbf{\Upsilon}_1^T,\ldots, \mathbf{\Upsilon}_K^T \right]^T$. Consequently, the overall SCM of the virtual array exhibits a block-partitioned structure given by
	\begin{equation}\label{16}
		\mathbf{R}=\frac{1}{N_p} \mathbf{\Upsilon} \mathbf{\Upsilon}^H=
		\begin{bmatrix}
			\mathbf{R}_{1,1} & \mathbf{R}_{1,2} & \cdots & \mathbf{R}_{1,K} \\
			\mathbf{R}_{2,1} & \mathbf{R}_{2,2} & \cdots & \mathbf{R}_{2,K} \\
			\vdots & \vdots & \ddots & \vdots \\
			\mathbf{R}_{K,1} & \mathbf{R}_{K,2} & \cdots & \mathbf{R}_{K,K}
		\end{bmatrix},
	\end{equation}
	where $\mathbf{R}_{k_1,k_2}=\mathbb{E}\{ \mathbf{\Upsilon}_{k_1} \mathbf{\Upsilon}_{k_2}^H\}\in \mathbb{C}^{N\times N}$ represents the cross-covariance sub-matrix between the $k_1$-th and $k_2$-th spatial sampling positions of the virtual array.

	%According to \eqref{1} and \eqref{3}, the received signal of the FA array without analog combining at the $k$-th spatial sampling position can be expressed as
	%\begin{equation}\label{15}
	%	\mathbf{\Upsilon}_k = \mathbf{H}_k\mathbf{S}+\mathbf{N}_k = \mathbf{A}_k\mathbf{\bar S}+\mathbf{N}_k
	%\end{equation}
	%the received signal of the large-scale virtual array with extended aperture can be constructed by $\mathbf{\Upsilon} = \left[ \mathbf{\Upsilon}_1^T,\ldots, \mathbf{\Upsilon}_K^T %\right]^T$, then the overall SCM of virtual array can be divided into
	%	\begin{equation}\label{16}
		%		\mathbf{R}=\frac{1}{N_p} \mathbf{\Upsilon} \mathbf{\Upsilon}^H=
		%	\begin{bmatrix}
			%		\mathbf{R}_{1,1} & \mathbf{R}_{1,2} & \cdots & \mathbf{R}_{1,K} \\
			%			\mathbf{R}_{2,1} & \mathbf{R}_{2,2} & \cdots & \mathbf{R}_{2,K} \\
			%			\vdots & \vdots & \ddots & \vdots \\
			%			\mathbf{R}_{K,1} & \mathbf{R}_{K,2} & \cdots & \mathbf{R}_{K,K}
			%		\end{bmatrix}
		%	\end{equation}
	%	where $\mathbf{R}_{k_1,k_2}=\mathbb{E}\{ \mathbf{\Upsilon}_{k_1} \mathbf{\Upsilon}_{k_2}^H\}\in \mathbb{C}^{N\times N}$ is the sub-SCM of virtual array.
	
	% The proposed method employs a stepwise reconstruction strategy: the FA array acquires multiple observation datasets through phase-shift adjustments at distinct spatial positions, followed by element-wise reconstruction of sub-SCM, which are subsequently synthesized into the global covariance matrix $\mathbf{R}$. The diagonal elements of the sub-SCM are prioritized in the reconstruction process, formulated as
	
	The proposed method employs a stepwise reconstruction strategy whereby the FA array acquires multiple observation datasets through systematic phase-shift configurations at distinct spatial positions, followed by element-wise reconstruction of the sub-SCMs, which are subsequently assembled into the global covariance matrix $\mathbf{R}$. 
	{In each sub-SCM, the diagonal elements are reconstructed with priority. Specifically, the $n$-th diagonal element is computed as}
	\begin{equation}\label{17}
		\mathbf{\hat R}_{k_1,k_2}(n,n) = N\mathbf{\hat w} _n^H \mathbf{R}_{k_1,k_2} \mathbf{\hat w} _n = R_{n,n},
	\end{equation}
	where $\hat{\mathbf{w}}_n=\frac{1}{\sqrt{N}}\mathbf{b}_n$ represents the reconstruction weight vector, and $\mathbf{b}_n\in\mathbb{C}^{N\times 1}$ is defined as a canonical basis vector with unity at the $n$-th position and zeros elsewhere. The physical implementation of this weight vector is achieved through selective antenna activation in the switching network, while setting the phase shift coefficients of the activated elements to zero.

	%where $\mathbf{\hat w}_n=\frac{1}{\sqrt{N}}\mathbf{b}_n$ represents the weight vector used at the corresponding position of the reconstruction, with the $\mathbf{b}_n\in\mathbb{C}^ {N\times 1}$ is defined as a canonical basis vector with unity at the $n$-th position and zeros elsewhere. The physical realization of the weight vector is achieved by switching the connectivity of the array elements and zeroing the phase shift coefficient of the active elements. In the design of \eqref{2}, the analog phase shifter satisfies the constant-modulus constraint. In order to eliminate the systematic reconstruction process for amplitude calibration, the error calibration factor $N$ is multiplied.
	
	%The remaining $N^2-N$ off-diagonal elements exist as conjugate-symmetric pairs. For each diagonally symmetric element pair  $\mathbf{ R}_{k_1,k_2}(n,m)$ and $\mathbf{ R}_{k_1,k_2}(m,n)$, we solve them jointly. Reconstructing this pair of elements requires two transformations of the analog combining vector vector. In the first adjustment, both the $n$-th and $m$-th FAs are activated with null phase-shift coefficients, resulting in a weight vector expressed as the sum of two canonical basis vector as $\mathbf{\bar w}_{n,m}=\frac{1}{\sqrt{N}}(\mathbf{b}_n+\mathbf{b}_m)$. Based on the weight vector design, we have
	
	The remaining $N^2-N$ off-diagonal elements exist as conjugate-symmetric pairs. For each conjugate-symmetric element pair $\mathbf{R}_{k_1,k_2}(n,m)$ and $\mathbf{R}_{k_1,k_2}(m,n)$, we solve them jointly through a two-measurement procedure. Reconstructing this pair of elements requires two distinct configurations of the analog combining vector.
	
	In the first measurement, both the $n$-th and $m$-th FAs are simultaneously activated with zero phase-shift coefficients, resulting in a weight vector expressed as the sum of two canonical basis vectors: $\mathbf{\bar{w}}_{n,m}=\frac{1}{\sqrt{N}}(\mathbf{b}_n+\mathbf{b}_m)$. Based on this weight vector design, the first measurement yields:  
	\begin{equation}\begin{aligned}\label{18}
			&\mathbf{\hat R}_{k_1,k_2}(n,m)  \\
			&= \hat{R}_{n,m} = N\mathbf{\bar w}_{n,m}^{H}\mathbf{R}_{k_1,k_2}\mathbf{\bar w}_{n,m}\\&=
			(\mathbf{b}_{n}+\mathbf{b}_{m})^{H}\mathbf{R}_{k_1,k_2}(\mathbf{b}_{n}+\mathbf{b}_{m}) \\
			& =\mathbf{R}_{k_1,k_2}(n,n)+\mathbf{R}_{k_1,k_2}(n,m)+\mathbf{R}_{k_1,k_2}(m,n) \\
			& \quad +\mathbf{R}_{k_1,k_2}(m,m).
	\end{aligned}\end{equation}
	%Consequently, for   the second weight vector adjustment, while preserving the physical connectivity between the $n$-th and $m$-th FAs, a randomly selected phase coefficient $\alpha\in(0,\pi/2)$  is applied, resulting in a weight vector formulated as the linear combination as $\mathbf{\bar w}_{m,n}=\frac{1}{N}(\mathbf{b}_{m}\mathbf{e}^{-j\alpha}+\mathbf{b}_{n}\mathbf{e}^{j\alpha})$, then we have 
	For the second measurement, both the $n$-th and $m$-th FAs remain activated while differential phase shifts are introduced. Specifically, a phase coefficient $\alpha$ is applied such that the $m$-th FA receives phase shift $-\alpha$ and the $n$-th FA receives phase shift $+\alpha$. This results in a weight vector expressed as the linear combination $\mathbf{\bar{w}}_{m,n}=\frac{1}{\sqrt{N}}(\mathbf{b}_{m}e^{-j\alpha}+\mathbf{b}_{n}e^{j\alpha})$. The second measurement then yields:
	\begin{equation}\begin{aligned}\label{19}
			&\mathbf{\hat R}_{k_1,k_2}(m,n)  \\
			&= \hat{R}_{m,n} =N\mathbf{\bar w}_{m,n}^{H}\mathbf{R}_{k_1,k_2}\mathbf{\bar w}_{m,n}
			\\&=(\mathbf{b}_{m}\mathbf{e}^{-j\alpha}+\mathbf{b}_{n}\mathbf{e}^{j\alpha})^{H}\mathbf{R}_{k_1,k_2}(\mathbf{b}_{m}\mathbf{e}^{-j\alpha}+\mathbf{b}_{n}\mathbf{e}^{j\alpha})\\
			& =\mathbf{R}_{k_1,k_2}(n,n)+\mathrm{e}^{-j2\alpha}\mathbf{R}_{k_1,k_2}(n,m) \\
			& \quad+\mathrm{e}^{j2\alpha}\mathbf{R}_{k_1,k_2}(m,n)+\mathbf{R}_{k_1,k_2}(m,m)
	\end{aligned}\end{equation}
	Since $\mathbf{\hat R}_{k_1,k_2}(n,n)$ and $\mathbf{\hat R}_{k_1,k_2}(m,m)$ has been obtained in advance,   \eqref{18} and \eqref{19} can be combined for joint solving via   
	
	\begin{equation}\label{20}
		\left.\left\{
		\begin{array}
			{l}\mathrm{e}^{-j2\alpha}\mathbf{R}_{k_1,k_2}(n,m)+\mathrm{e}^{j2\alpha}\mathbf{R}_{k_1,k_2}(m,n)=\widetilde{R}_{m,n} \\
			\mathbf{R}_{k_1,k_2}(n,m)+\mathbf{R}_{k_1,k_2}(m,n)=\widetilde{R}_{n,m}
		\end{array}\right.\right.\end{equation}
	Rewrite (19) into a more streamlined matrix form as
	\begin{equation}\label{21}
		\mathbf{E}\mathbf{r}_n=\mathbf{c}_n
	\end{equation}
	where $\mathbf{E}= \begin{bmatrix} \mathrm{e}^{-j2\alpha} & \mathrm{e}^{j2\alpha} \\ 1 & 1 \end{bmatrix}$, $\mathbf{r}_n= \begin{bmatrix} \mathbf{R}_{k_1,k_2}(n,m)  \\ \mathbf{R}_{k_1,k_2}(m,n)  \end{bmatrix}$ and $\mathbf{c}_n= \begin{bmatrix} \widetilde{R}_{m,n}  \\ \widetilde{R}_{n,m}  \end{bmatrix}$. By selecting $\alpha$ to satisfy $\mathrm{e}^{j2\alpha}\neq\mathrm{e}^{-j2\alpha}$,  $\mathbf{E}$  maintains full rank and invertibility. Consequently, the reconstructed cross-correlation terms in $ \mathbf{R}_{k_1,k_2}$ are analytically determined through the operation
	\begin{equation}\label{22}
		\left.\left\{
		\begin{array}
			{c}\hat{\mathbf{R}}_{k_1,k_2}(n,m)=\rho(\widetilde{R}_{m,n}-\mathrm{e}^{j2\alpha}\widetilde{R}_{n,m}) \\
			\hat{\mathbf{R}}_{k_1,k_2}(m,n)=\rho(\mathrm{e}^{-j2\alpha}\widetilde{R}_{n,m}-\widetilde{R}_{m,n})
		\end{array}\right.\right.
	\end{equation}
	where $\rho=1/(\mathrm{e}^{-j2\alpha}-\mathrm{e}^{j2\alpha})$.
	
	Through systematic traversal of indices $n,m\in[1,N]$ and iterative application of the reconstruction protocol defined in \eqref{17}-\eqref{22}, full-element reconstruction of $\mathbf{R}_{k_1,k_2}$ is achieved. Subsequently, distributed sub-SCMs are integrated via \eqref{16}, ultimately constructing a global covariance matrix that fully characterizes the spatial statistical properties of large-scale virtual arrays.  
	\footnote{Once the virtual-array SCM is reconstructed, the subsequent DOA estimation is carried out in a standard covariance model, which coincides with the classical fully-digital (virtual-array) setting. The corresponding CRLB has been extensively studied in the array processing literature and is largely agnostic to the hybrid front-end design parameters (e.g., the analog combiner and the phase-sampling budget). Therefore, we do not repeat this classical bound.}

	\subsection{JAD-RD-MUSIC Algorithm}
	The reconstructed SCM restores a higher-dimensional covariance representation, which improves subspace separability and, in turn, strengthens multi-source resolvability and DOA estimation accuracy. However, the resulting dimensionality expansion also implies a substantial increase in signal processing complexity.
	Moreover, the FA-enabled virtual array is generally irregular, so the associated manifold lacks the structured geometry exploited by fast uniform-array algorithms. 
	These considerations motivate the development of a geometry-agnostic yet computationally efficient 2-D DOA estimator, for which we next introduce a Jacobi-Anger expansion-based dimension-reduced MUSIC Algorithm. \footnote{Jacobi--Anger representations have also been explored for parameter estimation with arbitrary array geometries under hybrid front-end constraints (e.g.,~\cite{r42}). In contrast, here the expansion is used as an analytic tool to decouple the azimuth/elevation dependence in the MUSIC spectrum built upon the reconstructed virtual-array SCM, enabling an efficient reduced-dimension 2-D DOA search.}
	
	Jacobi-Anger expansion is a kind of polynomial expansion and generally given by 
	\begin{equation}\label{23}
		e^{jz\sin\theta}=\sum_{l=-\infty}^{\infty}J_{l}\left(z\right)e^{jl\theta}
	\end{equation}
	\begin{figure}[!t]
		\centering
		\includegraphics[width=3.6in]{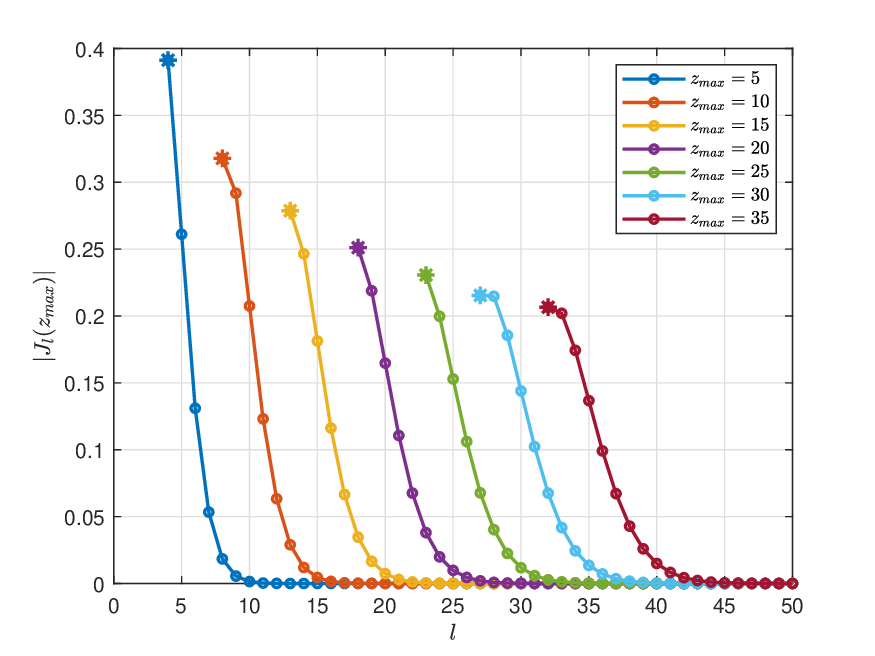}
		\caption{Bessel function of first kind versus different order $l$ (starting point of the curve is the global maximum of $|J_{l}(z_{\mathrm{max}})|$).}
		\label{Fig.2}
	\end{figure}
	where $J_{l}(z)$ is the Bessel function of first kind. Thus, the $\dot n$-th element in $\mathbf{a}(\theta, \phi)$ can be expressed as 
	\begin{equation}\label{24}
		\begin{aligned}
			a_{\dot n}\left(\theta,\phi\right) 
			& =e^{jz_{\dot n}(\phi)\sin\theta}
			=\sum_{l=-\infty}^{\infty}J_{l}\left(z_{\dot n}\left(\phi\right)\right)e^{jl\theta}
		\end{aligned}
	\end{equation}
	where $z_{\dot n}(\phi)=\frac{2\pi}{\lambda} (x_{\dot n}\cos\phi+y_{\dot n}\sin\phi)$ represents the term related to elevation angles $\phi$. Maintaining DOA estimation accuracy requires precise control of the Jacobi expansion order. As the order $l$ of the Jacobi-Anger expansion approaches the independent variable $z$, the amplitude of the first kind Bessel function $J_{l}(z)$ reaches its global maximum, followed by an exponential decay trend, as shown in Fig. \ref{Fig.2}. To maintain the required numerical accuracy of the series expansion throughout the entire computational domain, a truncation threshold is introduced to optimize the selection of expansion order
	\begin{equation}\label{25}
		|J_{l}(z)|<\epsilon
	\end{equation}
	where $\epsilon$ represents the predefined error tolerance threshold. This criterion ensures that the influence of higher-order terms on spectral estimation results becomes negligible. The choice of $\epsilon$ governs an explicit accuracy--complexity trade-off: a smaller $\epsilon$ yields a larger retained order $L_1$, which reduces the manifold approximation error and sharpens the MUSIC pseudospectrum peaks, but increases the dimension of $\mathbf{B}(\phi)$ and the associated computational cost; conversely, a larger $\epsilon$ reduces $L_1$ and computational burden at the expense of potential subspace leakage. In our implementation, $\epsilon=10^{-3}$ provides a practical balance. Let $L_1$ denote the minimum non-negative expansion order satisfying the constraint in (25). Then, $a_{\dot n}\left(\theta,\phi\right)$ can be approximated as
	\begin{equation}\label{26}
		\begin{aligned}
			a_{\dot n}\left(\theta,\phi\right)&\approx\sum_{l=-L_1}^{L_1}J_{l}\left(z_{\dot n}\left(\phi\right)\right)e^{jl\theta}\\&=\mathbf{b}(z_{\dot n}(\phi))\mathbf{e}(\theta)
		\end{aligned}
	\end{equation}
	where $\mathbf{b}(z_{\dot n}(\phi))=\left[J_{-L_1}\left(z_{\dot n}\left(\phi\right)\right),\cdots,J_{L_1}\left(z_{\dot n}\left(\phi\right)\right) \right] $ and $\mathbf{e}(\theta)=\left[e^{-jL_1\theta},\cdots,1,\cdots,e^{jL_1\theta} \right] $. The array manifold of the virtual array is given as
	\begin{equation}\label{27}
		\mathbf{a}(\theta,\phi)=\mathbf{B}(\phi)\mathbf{e}(\theta)
	\end{equation}
	where $\mathbf{B}(\phi)=\left[\mathbf{b}^T(z_1(\phi)),\cdots,\mathbf{b}^T(z_{\dot N}(\phi)) \right]\in \mathbb{C}^ {\dot N\times (2L_1 +1)}$ \footnote{The inherent column rank deficiency of the complex Jacobi-Anger expansion coefficient matrix $\mathbf{B}(z)$ arises from the odd symmetry property of the Jacobi functions $J_{-l}(z)=(-1)^{l}J_{l}(z)$.}.
	By leveraging the  polynomial expansion decomposition, the DOA estimation problem in \eqref{14} can be reformulated as
	\begin{equation}\label{28}
		\begin{aligned}
			\min_{\theta,\phi} \mathbf{e}^H(\theta) &\mathbf{B}^H(\phi) \mathbf{\bar U}_n \mathbf{\bar U}_n^H \mathbf{B}(\phi)\mathbf{e}(\theta)\\ &
			\mathrm{s.t.~}\mathbf{d}^{H}\mathbf{e}\left(\theta\right)=1
		\end{aligned}
	\end{equation}
	where $\mathbf{\bar U}_n$ denotes the noise subspace obtained from the EVD of the reconstructed SCM $\mathbf{\hat R}$ and the constraint vector $\mathbf{d}=\begin{bmatrix}0,\ldots,0, 1,0,\cdots,0\end{bmatrix}^{T}\in \mathbb{R}^{(2L_1+1) \times 1}$ is used to eliminate trivial all-zero solutions. Using the Lagrange multiplier method, we construct the Lagrangian from \eqref{28} as
	\begin{equation}\label{29}
		L(\theta, \phi) = \mathbf{e}^H(\theta) \mathbf{E}(\phi) \mathbf{e}(\theta) - \lambda \left( \mathbf{d}^{H} \mathbf{e}(\theta) - 1 \right)
	\end{equation}
	where $\mathbf{E}(\phi)=\mathbf{B}^H(\phi) \mathbf{\bar U}_n \mathbf{\bar U}_n^H \mathbf{B}(\phi)$ and $\lambda$ is the Lagrange multiplier.
	
	Setting the partial derivative of \eqref{29} with respect to $\mathbf{e}(\theta)$ to zero  yields $\mathbf{e}(\theta)=\mu\mathbf{E}^{\dagger}(\phi)\mathbf{d}$, where $\mu$ is a scaling constant. Incorporating the constraint $\mathbf{d}^{H}\mathbf{e}\left(\theta\right)=1$, we analytically determine  $\mu =1/\mathbf{d}^H\mathbf{E}^{\dagger}(\phi)\mathbf{d}$, ultimately obtaining the closed-form expression
	\begin{equation}\label{30}
		\mathbf{e}(\theta)=\frac{\mathbf{E}^{\dagger}(\phi)\mathbf{d}}{\mathbf{d}^H \mathbf{E}^{\dagger}(\phi) \mathbf{d}}
	\end{equation}
	
	Substituting \eqref{30} back into the spatial spectrum function  $P(\theta,\phi)$, the $L$ elevation angle estimates are computed through spectral peak search via
	\begin{equation}\label{31}
		\hat \phi = \mathop {\min }\limits_{\phi} \frac{1}{\mathbf{d}^H\mathbf{E}^{\dagger}(\phi)\mathbf{d}}
	\end{equation}
	
	The azimuth estimates are ultimately derived by substituting the elevation estimates from \eqref{31} into the original spectral function optimization framework:
	\begin{equation}\label{32}
		\hat \theta = \mathop {\max }\limits_{\theta} \frac{1}{{\bf{a}}^H(\theta ,\hat \phi ){{\bf \bar U}_{n}}{\bf \bar U}_n^H{\bf{a}}({\bf{\theta }},\hat \phi)}
	\end{equation}
	This procedure inherently guarantees automatic azimuth-elevation angle pairing. The implementation procedure of the 2-D DOA estimator based on the Jacobi-Anger decomposition (abbreviated as JAD-RD-MUSIC) is summarized in \textbf{Algorithm 2}.
	\begin{algorithm}[!t]
		\caption{JAD-RD-MUSIC for FA array}
		\LinesNumbered
		\KwIn{Reconstructed virtual array SCM $\mathbf{\hat R}$ and error tolerance $\epsilon$.}
		\KwOut{$(\hat \theta _\ell,\hat \phi _\ell) $}
		Subspace extraction:
		$\hat{\mathbf R}\xrightarrow{\mathrm{EVD}}\{\bar{\mathbf U}_s,\bar{\mathbf U}_n\}$.\\
		
		\For{ $\phi\in$ search grid}{
			Compute $z_{\dot n}(\phi) = \frac{2\pi}{\lambda}(x_{\dot n}\cos\phi + y_{\dot n}\sin\phi)$ for all $\dot n$, and let $z_{\max}=\max_{\dot n}|z_{\dot n}(\phi)|$.\\
			Find the smallest $L_1$ such that $|J_{L_1+1}(z_{\max})|<\epsilon$.\\
		    Form $\mathbf B(\phi)$ with $\mathbf b(z_{\dot n})=[J_{-L_1}(z_{\dot n}),\ldots,J_{L_1}(z_{\dot n})]$. \\
		$\mathbf E(\phi)\leftarrow \mathbf B^H(\phi)\bar{\mathbf U}_n\bar{\mathbf U}_n^H\mathbf B(\phi)$. \\
		Evaluate $S(\phi) = \mathbf d^H\mathbf E^\dagger(\phi)\mathbf d$. \\
		}
		Obtain $\{\hat\phi_\ell\}_{\ell=1}^L$ as the $L$ largest peaks of $S(\phi)$. \\
		{
			\For{ $\theta\in$ search grid}
			{
				Construct $\mathbf{a}(\theta,\hat{\phi}_\ell)$ by $\mathbf{a}(\theta,\phi) = [\mathbf{a}_1(\theta,\phi)^T,...,\mathbf{a}_K(\theta,\phi)^T]^T$.\\
				Update spectral function $P(\theta|\hat{\phi}_\ell) = 1/\|\mathbf{\bar U}_n^H\mathbf{a}(\theta,\hat{\phi}_\ell)\|^2$.\\
				Return $\hat{\theta}_\ell = \arg\max_{\theta} P(\theta|\hat{\phi}_\ell)$.\\
			}
		}
		Return 2-D DOA pairs $\{{(\hat \theta _\ell,\hat \phi _\ell)}\}_{\ell = 1}^L$.
	\end{algorithm}
	
	\emph{Remark 4:} The matrix
		$\mathbf{E}(\phi)=\mathbf{B}^H(\phi)\,\bar{\mathbf U}_n\bar{\mathbf U}_n^H\,\mathbf{B}(\phi)\in\mathbb{C}^{(2L_1+1)\times(2L_1+1)}$
		is inherently rank-deficient. In the reconstructed large-scale virtual-array regime, the projector $\bar{\mathbf U}_n\bar{\mathbf U}_n^H$ typically has rank well above $2L_1+1$, whereas the Jacobi--Anger coefficient matrix $\mathbf{B}(\phi)$ possesses an intrinsic column-rank deficiency; consequently, $\mathrm{rank}\{\mathbf{E}(\phi)\}<2L_1+1$ holds in general. Therefore, we employ the Moore-Penrose pseudoinverse $\mathbf{E}^\dagger(\phi)$ to ensure numerically stable inversion, while alternative regularizations (e.g., diagonal loading) can also be adopted. Moreover, the proposed Jacobi-Anger based peak-search replacement, when properly parameterized, can be readily incorporated into the FA-HAD-MUSIC procedure to further reduce the computational burden.

	\begin{table*}[!t]
		\centering 
		\caption{Comprehensive Comparison of DOA Estimation Methodologies for FAS 2-D DOA Estimation}
		\label{tab:doa_method_comparison}
		\begin{tabular}{|l|c|c|l|c|}
			\hline
			\multirow{2}{*}{\textbf{Characteristics}} & \multicolumn{2}{c|}{\textbf{Method Comparison}} & \multirow{2}{*}{\textbf{Performance Metrics}} & \multirow{2}{*}{\textbf{Trade-off Analysis}} \\
			\cline{2-3}
			& \textbf{FA-HAD-MUSIC} & \textbf{SCM + JAD-RD-MUSIC} & & \\
			\hline
			\textbf{Data Processing} & Direct compressed & Systematic full & Complexity vs. & FA-HAD: Speed \\
			\textbf{Strategy} & observations & reconstruction & Accuracy & SCM: Precision \\
			\hline
			\textbf{Covariance Matrix} & $KT \times KT$ & $NK \times NK$ & Matrix size & $T \ll N$: FA-HAD wins \\
			\textbf{Dimension} & (compressed) & (full virtual array) & impacts EVD cost & $T \approx N$: SCM wins \\
			\hline
			\textbf{Sampling} & Random phase shifts & Two-step measurement & SNR preservation & FA-HAD: $-3$dB penalty \\
			\textbf{Mechanism} & ($T$ per position) & protocol & vs. efficiency & SCM: No SNR loss \\
			\hline
			\textbf{Search Strategy} & 2-D joint search & Decoupled elevation- & Search efficiency & Joint: $n_\theta \times n_\phi$ \\
			& $(\theta, \phi)$ & azimuth approach & & Decoupled: $n_\theta + n_\phi$ \\
			\hline
			\textbf{Target Resolution} & Limited by $KT$ & Enhanced by $NK$ & Maximum detectable & FA-HAD: $\leq KT-1$ \\
			\textbf{Capability} & signal dimensions & virtual aperture & targets & SCM: $\leq NK-1$ \\
			\hline
			\textbf{Measurement} & $TK$ pilots & $N^2K$ & Pilot overhead & When $T=3, N=8$: \\
			\textbf{Overhead} & & measurements & & FA-HAD: $3K$, SCM: $64K$ \\
			\hline
			\textbf{Computational} & $\mathcal{O}((KT)^3)$ & $\mathcal{O}((NK)^3)$ & Processing time & FA-HAD: Real-time \\
			\textbf{Complexity} & $+ (NK)^2n_\theta n_\phi$ & $+ \bar{L}_1(NK)^2n_\theta$ & scalability & SCM: Near real-time \\
			\hline
			\textbf{Robustness} & Moderate (phase & High (deterministic & Noise resilience & Low SNR: SCM better \\
			& randomization) & reconstruction) & & High SNR: Comparable \\
			\hline
			\textbf{Array Geometry} & Regular/irregular & Universal & Flexibility & FA-HAD: Configuration \\
			\textbf{Adaptability} & configurations & (arbitrary topology) & requirements & SCM: Topology-agnostic \\
			\hline
		\end{tabular}
	\end{table*}
	
	\emph{Remark 5:} Table \ref{tab:doa_method_comparison} presents a comprehensive analysis of the trade-offs between the two proposed DOA estimation methodologies. The FA-HAD-MUSIC algorithm prioritizes computational efficiency and real-time performance, making it ideal for applications requiring rapid response with moderate accuracy requirements. In contrast, the SCM reconstruction combined with JAD-RD-MUSIC approach emphasizes estimation precision and robustness, particularly excelling in challenging low SNR scenarios and multi-target environments. The choice between these methodologies should be guided by specific application constraints: for time-critical applications with sufficiently high SNR, FA-HAD-MUSIC provides optimal efficiency; for high-precision sensing applications or harsh propagation conditions, the SCM-based approach delivers superior performance despite increased computational overhead.
	
	\subsection{Computational Complexity}
	In subspace-based methods, the dominant computational costs correspond to three operations: calculation of the covariance matrix, its associated EVD, and subsequent spatial spectrum scanning. For an FA array system employing $N$ elements, the FA-HAD-MUSIC algorithm based on an HAD architecture under $K$ movements with $T$ observation measurements per movement exhibits the following complexity:  the SCM computation requires $\mathcal{O}((KT)^2N_p)$ operations, the EVD consumes $\mathcal{O}((KT)^3)$ operations, and the 2-D spatial spectrum search over an ${n_\theta}\times {n_\phi}$ grid introduces $\mathcal{O}((NK)^2{n_\theta}{n_\phi})$ operations, yielding an overall computational complexity of $\mathcal{O}\big((KT)^2N_p +(KT)^3 + (NK)^2{n_\theta}{n_\phi}\big)$. Compared to fully digital FA arrays with identical configurations, the proposed FA-HAD-MUSIC method significantly reduces computational overhead in SCM computation and EVD through effective data compression. In contrast, the JAD-RD-MUSIC algorithm is originally designed for fully digital FA arrays with arbitrary virtual array geometries. Its computational complexity is characterized by: $\mathcal{O}((NK)^2N_p)$ for SCM computation, $\mathcal{O}((NK)^3)$ operations for EVD and $\mathcal{O}((\bar{L}_1(NK)^2+{\bar{L}_1}^3){n_\theta} + L(NK)^2{n_\phi})$ operations for the dimensionality-reduced search process, resulting in a total complexity of  $\mathcal{O}\big((NK)^2N_p +(NK)^3 + (\bar{L}_1(NK)^2+{\bar{L}_1}^3){n_\theta} + L(NK)^2{n_\phi}\big)$, where $\bar{L}_1=2L_1+1$ represents the total number of dominant terms retained in the Jacobi-Anger expansion. It is worth noting that in practical systems, the number of single-point observations $T$ is typically much smaller than the number of antennas $N$.  This characteristic leads to significant differences in computational efficiency between the two algorithms across various application scenarios.
	
	\section{Simulation Results}
	In this section, the DOA estimation performance of the proposed algorithms is evaluated. Unless otherwise specified, the simulation parameters are configured as follows: The experimental setup consists of an $8$-element fluid antenna array, initialized as a uniform linear array (ULA) deployed along the $y$-axis. The array then performs $24$ position updates by moving along the $x$-axis, thereby forming an effective 2-D virtual aperture (equivalent planar sampling) for joint azimuth--elevation estimation. To enhance generality, a random movement pattern is adopted, where each step size is independently drawn from $[0.15\lambda,\,0.45\lambda]$. Constraining the step size to be within half a wavelength helps prevent spatial aliasing  while maintaining a high spatial DOFs utilization efficiency in the constrained region; meanwhile, randomized steps introduce geometric diversity and serve as a robustness test. The number of pilot transmissions (phase observations) $T$ is set to $1$, $3$, or $5$ in different cases. 
	We consider $L=6$ independent sensing signals under LOS propagation, with complex path gains $\gamma_\ell\sim\mathcal{CN}(0,1)$ and both azimuth and elevation angles uniformly distributed over $[-90^\circ,90^\circ]$. The Jacobi--Anger expansion uses a truncation threshold $\epsilon=10^{-3}$, and the SCM reconstruction employs $\alpha=\pi/8$. Results are averaged over $500$ Monte Carlo trials. Performance is evaluated using two metrics: the normalized squared error (NSE) for SCM reconstruction, $\mathrm{NSE}=\|\mathbf{R}-\hat{\mathbf{R}}\|_{F}^{2}\cdot\|\mathbf{R}\|_{F}^{-2}$, and the DOA root mean square error (RMSE), $\mathrm{RMSE}=\frac{1}{L}\sum_{\ell=1}^{L}\sqrt{\mathbb{E}\!\left[\big((\hat{\theta}_{\ell}-\theta_{\ell})^{2}+(\hat{\phi}_{\ell}-\phi_{\ell})^{2}\big)/2\right]}$.

\begin{figure}[!t]
		\centering
		\includegraphics[width=3.8in]{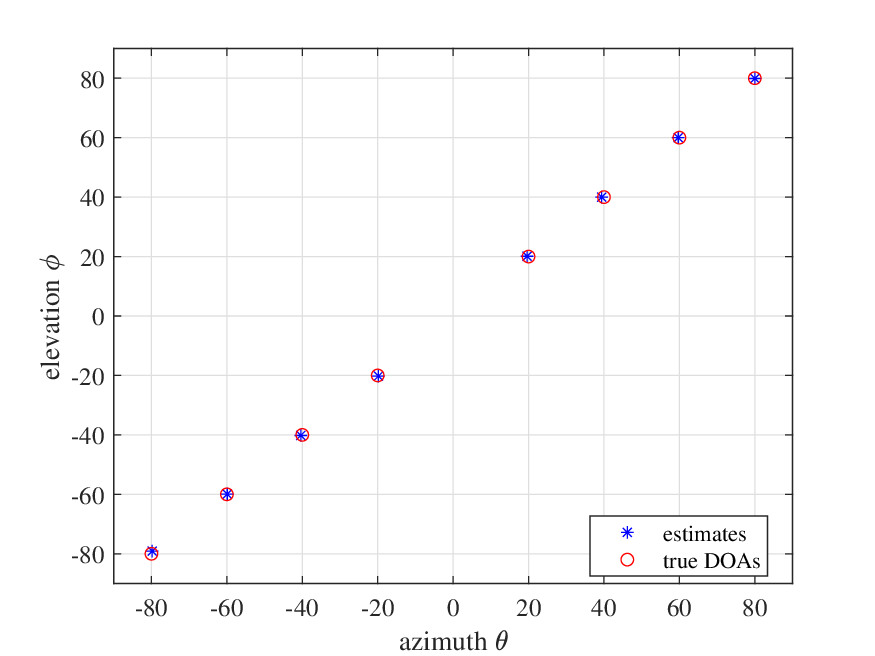}
		\caption{Comparison between DOA estimation results and ground truth.}
		\label{Fig.3}
	\end{figure}

	In the first simulation, to simplify the system model, we adopted a ULA with half-wavelength inter-element spacing as the initial array configuration. Only a single pilot signal was used for measurement at each antenna element position, with the received SNR fixed at 0 dB. As demonstrated in Fig. \ref{Fig.3}, the proposed strategy of incorporating random phase shifts to capture spatial angle information, combined with the fluid antenna design, enables highly accurate DOA estimation with minimal pilot overhead and hardware complexity.
	
	\emph{Remark 6:} The tight clustering of estimation points around the diagonal line in Fig. \ref{Fig.3} validates the fundamental effectiveness of the FA-HAD architecture. Despite operating with minimal resources ($T=1$ pilot per position), the algorithm achieves estimation errors consistently below 2 degrees across all {eight targets}, demonstrating that the coordinated spatial-phase sampling mechanism successfully captures the essential angular information without requiring exhaustive spatial scanning.
	
\begin{figure}[!t]
		\centering
		\includegraphics[width=3.8in]{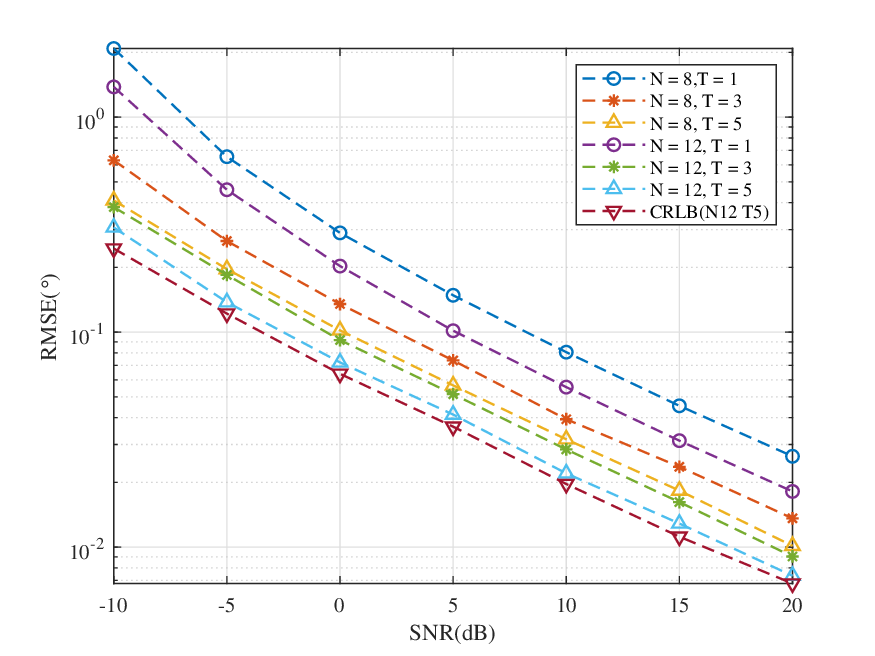}
		\caption{RMSE of DOA estimation versus SNR, $N$ and $T$, with $K=24$.}
		\label{Fig.4}
	\end{figure}

    In the second set of simulations, we examine how the initial array size $N$ and the number of phase observations $T$ affect DOA estimation. As shown in Fig.~\ref{Fig.4}, for a fixed $N$, increasing $T$ consistently reduces the RMSE, with the most visible gain appearing in the low-SNR region; meanwhile, the improvement gradually saturates as $T$ becomes larger, which agrees with the CRLB analysis indicating an approximately linear information accumulation with respect to $T$. When four additional elements are appended along the $y$-axis, the $N=12$ configuration consistently attains a lower RMSE than $N=8$ under the same motion pattern and the same $T$. This empirical advantage is expected: a larger physical aperture and richer spatial sampling improve subspace separability and multi-source resolvability. In this sense, increasing the initial array size enhances the efficiency of spatial-DOFs acquisition. Finally, the curves show that $(N=8,\,T=5)$ and $(N=12,\,T=3)$ yield comparable performance, illustrating a practical accuracy-overhead trade-off between pilot usage (via $T$) and array aperture/hardware scale (via $N$).

\begin{figure}[!t]
		\centering
		\includegraphics[width=3.8in]{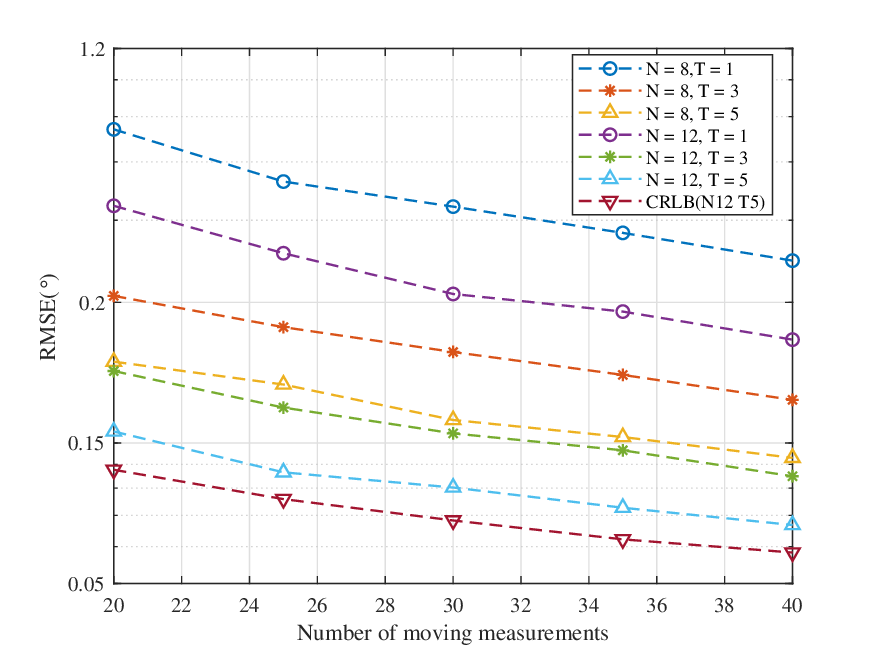}
		\caption{RMSE of DOA estimation versus $K$, $N$ and $T$, with SNR = 0 dB.}
		\label{Fig.5}
	\end{figure}

	The third set of simulations evaluates the effect of the number of spatial sampling positions (i.e., array movements) $K$ on DOA estimation performance. Fig.~\ref{Fig.5} reports the RMSE versus $K$ under the same set of $(N,T)$ configurations as in Fig.~\ref{Fig.4}. A clear and consistent trend is observed: for all curves, the RMSE decreases monotonically as $K$ increases from $20$ to $40$, confirming that additional movements provide more effective spatial sampling and enlarge the virtual aperture, thereby improving angle resolvability. Moreover, in the considered range, the performance gain brought by increasing $K$ remains pronounced, in contrast to the earlier saturation behavior observed when increasing $T$ beyond moderate values. This is consistent with the CRLB-based interpretation that $K$ not only accumulates information through more samples but also strengthens the geometry-induced spread of the virtual elements. Overall, the results further corroborate that $K$, together with $T$ and $N$, constitutes the key design knobs governing the accuracy-overhead trade-off of the proposed FA-HAD framework.
	
\begin{figure}[!t]
		\centering
		\includegraphics[width=3.8in]{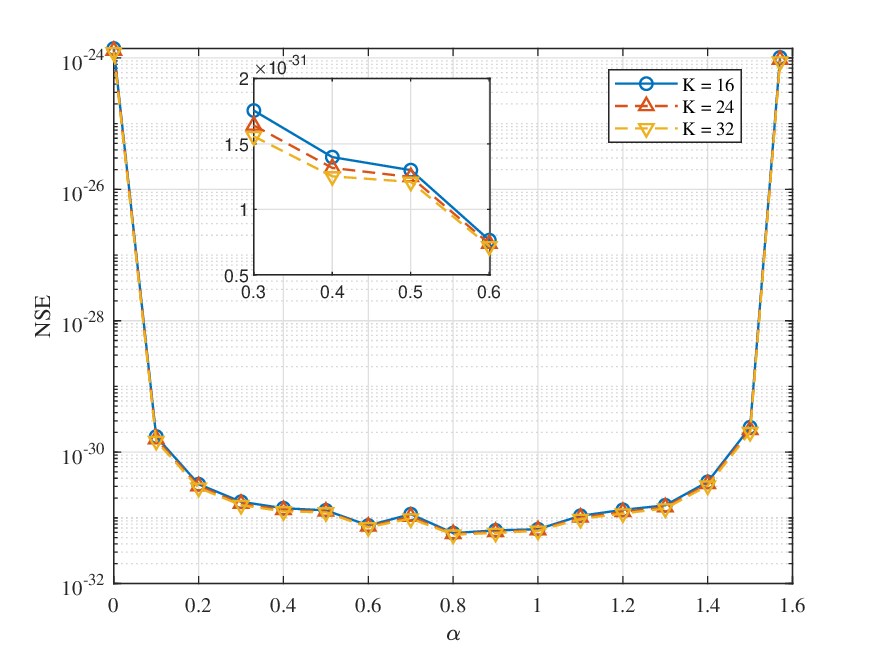}
		\caption{NSE of SCM reconstruction versus $\alpha$ and $K$.}
		\label{Fig.6}
	\end{figure}

	The fourth simulation verifies the impact of parameter $\alpha$ on the reconstruction accuracy of the virtual array SCM. The array consists of 8 FAs, with the number of movements varying among 16, 24, and 32. As shown in Fig. \ref{Fig.6}, the reconstruction accuracy remains high across most parameter ranges except when $\alpha$ approaches the extreme values of 0 and $\pi/2$. This occurs because $\mathbf{E}$ in \eqref{21} becomes ill-conditioned and the coefficient $\rho$ tends to infinity in these cases, leading to numerical instability. Hence, in practical applications, it is sufficient to avoid selecting parameter values near these critical regions. Furthermore, the NSE remains consistent across different movement counts. The reconstruction accuracy does not decrease with increased movements, demonstrating robustness of the proposed method.
	
	\emph{Remark 7:} The flat NSE curves in Fig. \ref{Fig.6} across different movement counts confirm the scalability robustness of the SCM reconstruction algorithm. The consistent performance regardless of $K$ values indicates that the two-step measurement protocol maintains numerical stability as the virtual array expands, {while the allowable range of the phase control parameter $\alpha$} provides sufficient operational flexibility for practical implementations.
	\begin{figure}[!t]
		\centering
		\subfloat[]{\includegraphics[width=3.8in]{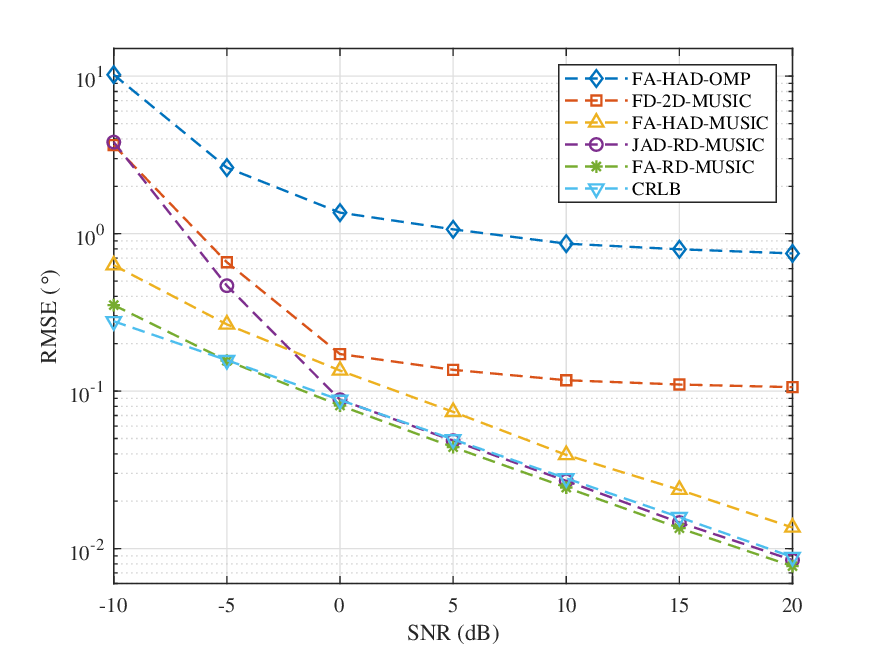}
			\label{subfig:alg-SNR}}
		\hfil
		\subfloat[]{\includegraphics[width=3.8in]{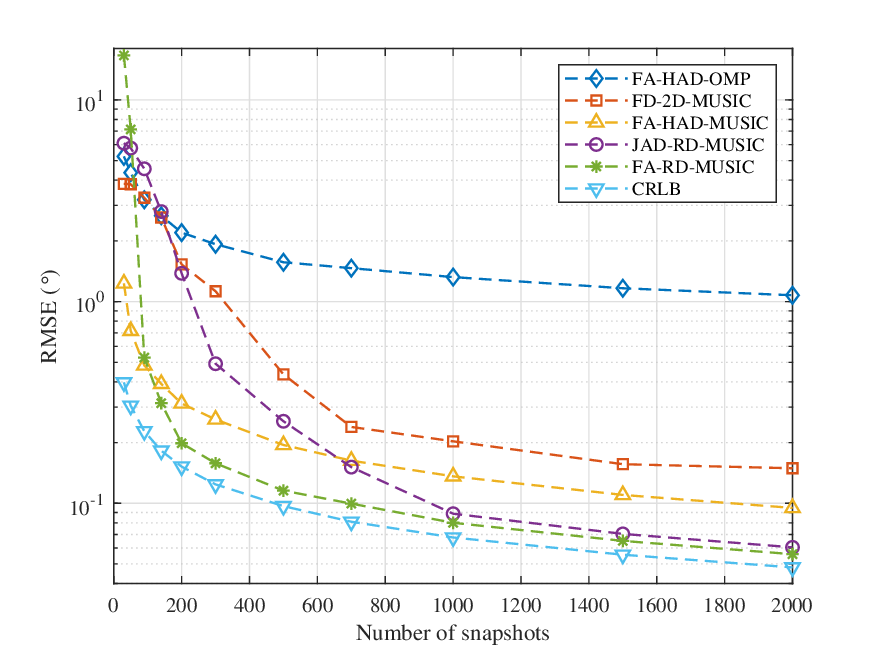}
			\label{subfig:alg-snap}}
		\caption{Comparisons of accuracy between different methods. 
			(a) RMSE versus SNR. (b) RMSE versus the number of snapshots.}
		\label{Fig.7}
	\end{figure}
	
	In the {fifth simulation}, we evaluate the DOA estimation performance of the proposed FA-HAD-MUSIC (with $T=3$) and JAD-RD-MUSIC algorithms. To ensure the rigor of comparative experiments, two representative benchmark algorithms are selected: One approach utilizes a fully digital uniform planar array (UPA) with half-wavelength inter-element spacing deployed under identical spatial constraints, integrated with the conventional 2-D-MUSIC algorithm (abbreviated as FD-2D-MUSIC) for angle estimation, while the other is based on reconstructed sample covariance matrices and employs the reduced-dimension MUSIC algorithm (denoted as FA-RD-MUSIC) in \cite{r37}. Moreover, we include OMP-based DOA estimator \cite{r38} as a representative sparse-recovery baseline tailored to the compressive observation setting (denoted as FA-HAD-OMP). We also include the CRLB evaluated for the corresponding FD (uncompressed) virtual-array model as an oracle baseline, since the SCM-reconstruction-based pipeline aims to recover this covariance-domain representation and thus the resulting CRB provides a consistent benchmark of the achievable accuracy.
	As shown in Fig. \ref{Fig.7}, the FA-HAD-MUSIC and JAD-RD-MUSIC algorithms demonstrate estimation performance comparable to fully digital architectures while exhibiting superior multi-source resolution capability. Notably, compared with FA-HAD-OMP, which also performs DOA estimation directly from compressive observations as a sparse-recovery baseline, the proposed FA-HAD-MUSIC achieves a more pronounced accuracy advantage while still maintaining stable and reliable estimation performance under reduced hardware complexity and pilot overhead. Benefiting from the innovative hybrid analog-digital architecture design, the system significantly reduces hardware implementation complexity without compromising performance. Furthermore, although the FA-RD-MUSIC algorithm shows certain advantages under low SNR conditions, the JAD-RD-MUSIC algorithm demonstrates superior adaptability due to its excellent architectural compatibility, effectively meeting the requirements of various FAS application scenarios.
	
	{\emph{Remark 8:} The performance convergence in Fig. \ref{Fig.7} between the proposed FA-HAD methods and fully digital benchmarks validates the architectural effectiveness across diverse operating conditions. Notably, the JAD-RD-MUSIC algorithm demonstrates superior snapshot efficiency, requiring fewer pilot symbols to achieve comparable accuracy, while FA-HAD-MUSIC excels in computational simplicity, making it ideal for real-time applications with stringent latency constraints.}
	
	\begin{figure}[!t]
		\centering
		\includegraphics[width=3.8in]{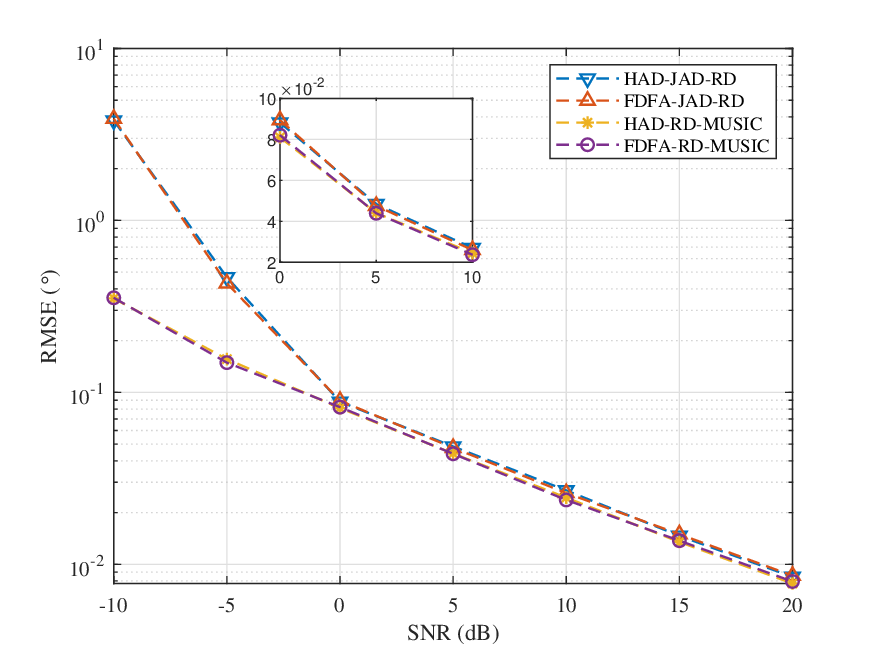}
		\caption{Architecture comparison: HAD vs. {fully digital FA (FDFA) array} under identical virtual array configurations and estimation algorithms.}
		\label{Fig.8}
	\end{figure}
	
	The sixth simulation investigates a fundamental architectural question in FAS: whether the analog compression inherent in HAD structures introduces performance degradation compared to fully digital implementations. This comparison is particularly critical given the substantial hardware complexity reduction achieved by HAD architectures. As illustrated in Fig. \ref{Fig.8}, under identical virtual array configurations and estimation algorithms, the proposed HAD architecture demonstrates performance parity with its fully digital counterpart {(FDFA)} across the entire SNR range. This remarkable consistency validates two key technical insights: First, the proposed SCM reconstruction methodology effectively compensates for the information loss introduced by analog combining, ensuring that the spatial diversity captured by the virtual array is fully preserved in the compressed domain. Second, the random phase sampling strategy, despite introducing a certain SNR penalty due to non-directional energy spreading, maintains sufficient signal diversity to support high-precision DOA estimation. The convergence of performance curves reveals that the fundamental limitation lies not in the analog processing chain, but rather in the finite spatial sampling resolution determined by the movement pattern $K$ and initial array size $N$. This finding has profound implications for system design: the dramatic hardware simplification achieved through single RF chain implementation (compared to $N$ RF chains in {FDFA}) comes at virtually no performance cost, enabling cost-effective deployment of multiple FA arrays for enhanced spatial coverage and sampling efficiency.
	
	\begin{figure}[!t]
		\centering
		\includegraphics[width=3.8in]{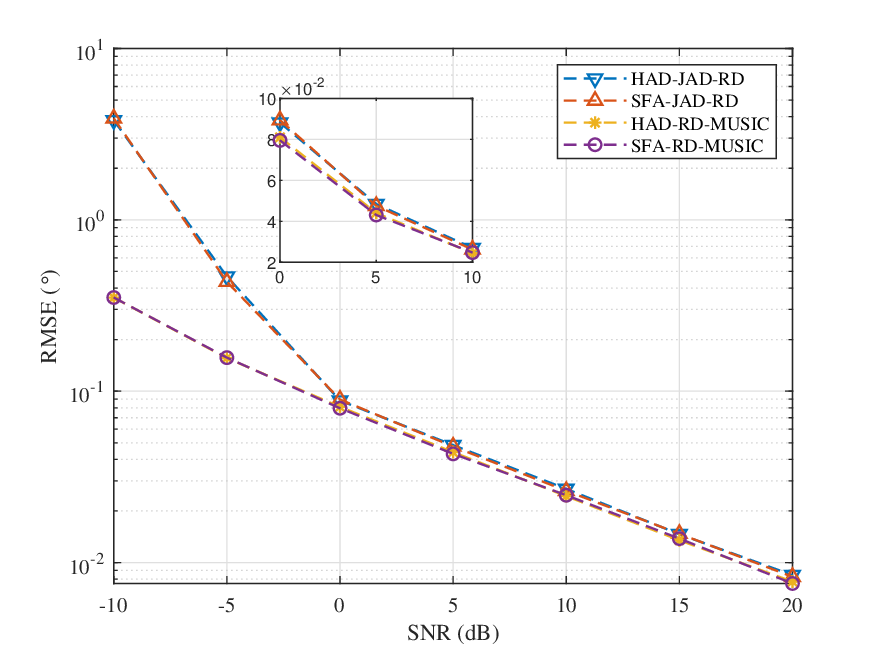}
		\caption{Mobility paradigm comparison: {Single FA (SFA)} sequential sampling vs. FA array coordinated movement under equivalent spatial coverage.}
		\label{Fig.9}
	\end{figure}
	
	The seventh simulation addresses a critical design choice in FAS implementation: whether to employ a single antenna with extensive mobility or a coordinated antenna array with limited displacement. This comparison directly impacts both system complexity and operational efficiency. As demonstrated in Fig. \ref{Fig.9}, the estimation accuracy of the proposed FA array approach closely matches that of single fluid antenna (SFA) systems when both methods sample equivalent spatial positions. However, this performance parity masks significant operational advantages that become apparent through deeper analysis. The FA array approach achieves identical spatial coverage through $K$ coordinated movements, while the SFA paradigm requires $NK$ individual positioning operations to sample the same virtual aperture. This represents an order-of-magnitude reduction in mechanical complexity and temporal overhead. Furthermore, the array-level mobility scheme exhibits superior resilience to positioning errors: small perturbations in individual element positions are averaged across the entire array, whereas SFA systems suffer from accumulative positioning errors over extended sampling sequences. The practical implications extend beyond computational efficiency to power consumption and system reliability. The coordinated movement mechanism reduces the number of nanopump activation cycles by a factor of $N$, directly translating to extended operational lifetime and reduced maintenance requirements. Additionally, the parallel spatial sampling capability enables adaptive real-time reconfiguration, where subsets of the array can be dynamically repositioned based on environmental feedback while maintaining continuous sensing coverage.
	
	\emph{Remark 9:} The comparative analysis presented in Figs. \ref{Fig.8} and \ref{Fig.9} establishes two fundamental principles for FAS design. First, the architectural choice between HAD and fully digital implementations should be driven primarily by cost and complexity constraints rather than performance considerations, as our proposed SCM reconstruction technique ensures performance equivalence. This finding challenges the conventional wisdom that analog processing necessarily compromises estimation accuracy, demonstrating instead that intelligent signal processing can overcome hardware-imposed limitations. Second, the mobility paradigm comparison reveals that coordinated array movement represents the optimal balance between spatial sampling efficiency and implementation complexity. While single antenna systems offer theoretical flexibility, the practical advantages of coordinated movement—including reduced mechanical overhead, enhanced positioning robustness, and improved power efficiency—make array-level mobility the preferred approach for real-world deployments. These insights collectively suggest that the proposed FA-HAD architecture represents a matured design paradigm that successfully bridges the gap between theoretical performance potential and practical implementation constraints, positioning fluid antenna technology for widespread adoption in next-generation wireless sensing and communication systems.
	
	Finally, to provide an intuitive assessment of the practical computational burden, we report the median CPU runtime (in seconds) per DOA estimation over 30 independent trials under the same settings as Simulation~1 (MATLAB R2020b, identical workstation). The results are summarized in Table~\ref{tab:runtime}. Consistent with the analytical complexity analysis in Section~III-C, FA-HAD-MUSIC achieves the shortest runtime owing to its reduced effective observation dimension, while JAD-RD-MUSIC offers a favorable trade-off between generality (arbitrary array geometry) and computational cost.
	
	\begin{table}[!t]
	\centering
	\caption{Comparison of CPU Runtime (s)}
	\label{tab:runtime}
	\setlength{\tabcolsep}{3.5pt}
	\renewcommand{\arraystretch}{1}
	\begin{tabular}{@{}c c c c@{}}
		\toprule
		\textbf{FD-2D-MUSIC} & \textbf{FA-HAD-MUSIC} & \textbf{JAD-RD-MUSIC} & \textbf{FA-RD-MUSIC} \\
		\midrule
		3.7737 & 0.5300 & 1.3491 & 0.7580 \\
		\bottomrule
	\end{tabular}
\end{table}

	\section{Conclusion}
	This paper proposed a novel FA-HAD architecture to more effectively exploit spatial degrees of freedom for high-precision 2-D DOA estimation under practical front-end constraints. A collaborative spatial-phase sampling strategy was developed to enable fast compressive-domain angle estimation, and its achievable accuracy was quantitatively characterized via a single-source CRLB analysis, providing principled guidance for the accuracy-overhead trade-off. To strengthen scalability and compatibility with broader array design frameworks, we further introduced an efficient virtual-array SCM reconstruction method that recovers a physically meaningful covariance representation and offers a reusable covariance-domain interface. Building upon the reconstructed SCM, a universal and computationally efficient 2-D DOA estimator was developed via the Jacobi-Anger expansion, supporting arbitrary planar  array geometries with a favorable complexity profile. Extensive simulations demonstrated that the proposed FA-HAD framework attains DOA accuracy close to FD baselines while substantially reducing RF hardware and training overhead. Future work will investigate joint combiner/mobility optimization, robustness under faster channel variations, and the impact of practical hardware non-idealities.
	
	\section*{Appendix}
	
	For the scenario of single signal source, we have  $\mathbf{\tilde{A}}=\mathbf{\tilde{a}}$, $\mathbf{\tilde{a}}=\mathbf{Q}^{-\frac{1}{2}}\pmb{\Phi}\mathbf{a}$, $\mathbf{\tilde{B}}=[ \mathbf{\tilde{b}}_1,\mathbf{\tilde{b}}_2] $, $\mathbf{\tilde{b}}_1=\mathbf{Q}^{-\frac{1}{2}}\pmb{\Phi}\mathbf{b}_1$, $\mathbf{\tilde{b}}_2=\mathbf{Q}^{-\frac{1}{2}}\pmb{\Phi}\mathbf{b}_2$, $\mathbf{b}_1=\frac{\partial \mathbf{a}}{\partial \theta}= j\frac{2\pi}{\lambda}  \cos\theta \mathbf{D}_1 \mathbf{a}$, $\mathbf{b}_2=\frac{\partial \mathbf{a}}{\partial \phi}=j\frac{2\pi}{\lambda} \sin\theta\mathbf{D}_2 \mathbf{a}$, and
	\begin{equation}\label{52}
		\mathbf{D}_1 = \mathrm{diag}\{\bar d_1,\cdots,\bar d_{\dot N}\},
	\end{equation}
\begin{equation}\label{53}
	\mathbf{D}_2 = \mathrm{diag}\{\tilde d_1,\cdots,\tilde d_{\dot N}\},
\end{equation}
	with $\bar d_{\dot n}=\left( x_{\dot n} \cos\phi + y_{\dot n} \sin\phi\right)  $ and $\tilde d_{\dot n}=\left( -x_{\dot n} \sin\phi + y_{\dot n} \cos\phi\right)$,\footnote{The equivalent virtual array model consists of $\dot N = NK$ virtual elements, with the 3D coordinates of the $\dot n$-th element defined as $(x_{\dot n},y_{\dot n},0)$.} which correspond to the $\dot n$-th diagonal elements of $\mathbf{D}_1$ and $\mathbf{D}_2$.
	
	Then \eqref{76} is rewritten as

	\begin{equation}\label{55}
		\mathbf{F}_{\pmb{\theta}\pmb{\theta}} = \frac{2N_p \hat{p} \left( \|\tilde{\mathbf{a}}\|_2^2 \|\tilde{\mathbf{b}}_1\|_2^2 - \left| \tilde{\mathbf{b}}_1^H \tilde{\mathbf{a}} \right|^2 \right)}{\sigma_n^2 \|\tilde{\mathbf{a}}\|_2^2},
	\end{equation}
	\begin{equation}\label{56}
		\mathbf{F}_{\pmb{\phi}\pmb{\phi}} = \frac{2N_p \hat{p} \left( \|\tilde{\mathbf{a}}\|_2^2 \|\tilde{\mathbf{b}}_2\|_2^2 - \left| \tilde{\mathbf{b}}_2^H \tilde{\mathbf{a}} \right|^2 \right)}{\sigma_n^2 \|\tilde{\mathbf{a}}\|_2^2},
	\end{equation}
	\begin{equation}\label{57}
		\mathbf{F}_{\pmb{\theta}\pmb{\phi}} = \frac{2N_p \hat{p} \left( \|\tilde{\mathbf{a}}\|_2^2 \tilde{\mathbf{b}}_1^H \tilde{\mathbf{b}}_2-\tilde{\mathbf{b}}_1^H\tilde{\mathbf{a}}\tilde{\mathbf{a}}^H\tilde{\mathbf{b}}_2 \right)}{\sigma_n^2 \|\tilde{\mathbf{a}}\|_2^2},
	\end{equation}
	\begin{equation}\label{58}
		\mathbf{F}_{\pmb{\phi}\pmb{\theta}} = \frac{2N_p \hat{p} \left( \|\tilde{\mathbf{a}}\|_2^2 \tilde{\mathbf{b}}_2^H \tilde{\mathbf{b}}_1-\tilde{\mathbf{b}}_2^H\tilde{\mathbf{a}}\tilde{\mathbf{a}}^H\tilde{\mathbf{b}}_1 \right)}{\sigma_n^2 \|\tilde{\mathbf{a}}\|_2^2},
	\end{equation}
	with $\hat{p} = \frac{1}{N_p } \sum_{n=1}^{N_p } | \bar s(n)|^2$ being the estimate of $p = \mathbb{E} \{ |\bar s(n)|^2 \}$. 
	
	Under the assumption $\mathbf{Q}\approx\mathbf{I}$,\footnote{This approximation is reasonable for two reasons. First, due to the modular-based selection, $\pmb{\Phi}$ has a regular block-wise support pattern, and thus $\mathbf{Q}$ exhibits a structured (near block-diagonal) form dominated by its diagonal blocks. Second, with random phase shifts, the cross-terms correspond to products of random phasors whose cross-correlation is much weaker than the auto-correlation, rendering the off-diagonal entries negligible.} we calculate $\|\tilde{\mathbf{a}}\|_2^2$,  $\|\tilde{\mathbf{b}}_1\|_2^2$ and $\left| \tilde{\mathbf{b}}_1^H \tilde{\mathbf{a}} \right|^2$ separately for $\mathbf{F}_{\pmb{\theta}\pmb{\theta}}$. First, 
	$\|\tilde{\mathbf{a}}\|_2^2$ is computed by
	\begin{equation}\label{59}
		\begin{gathered}
			\left\|\tilde{\mathbf{a}}\right\|_{2}^{2}= \mathbf{a}^H\mathbf{\Phi}^H \mathbf{\Phi}\mathbf{a}.
		\end{gathered}
	\end{equation}
	
	Let $\Omega(k)$ denote the set of positions of non-zero elements in the $k$-th row of matrix $\mathbf{\Phi}$. Then, we can express the $k$-th element of $\mathbf{\Phi}\mathbf{a}$ as $ \frac{1}{\sqrt{N}} \sum_{\dot n \in \Omega(k)} e^{j\left(w_{\dot n}-\alpha_{k,\dot n}\right)} $,  which yields
	\begin{equation}\label{60}
		\begin{aligned}
			&\left\|\tilde{\mathbf{a}}\right\|_{2}^{2}= \frac{1}{N}\sum_{k=1}^{KT}  \Big(\sum_{\dot n \in \Omega(k)}e^{j\left(\alpha_{k,\dot n}-w_{\dot n}\right)}\sum_{\dot n^{\prime}\in \Omega(k)}e^{j\left(w_{\dot n'}-\alpha_{k,\dot n^{\prime}}\right)}\Big) \\
			&=K T+\frac{1}{ N}\sum_{k=1}^{K T}\sum_{\dot n \in \Omega(k)}\sum_{\dot n^{\prime}\neq \dot n}e^{j(\alpha_{k,\dot n}-w_{\dot n})+j(w_{\dot n'}-\alpha_{k,\dot n^{\prime}})},
		\end{aligned}
	\end{equation}
	The second term is zero due to property that the characteristic function of $e^{j\alpha}$ with $\alpha\sim\mathcal{U}(0, 2\pi)$ is zero.
	
	The term $\|\tilde{\mathbf{b}}_1\|_2^2$ is:
	\begin{equation}\label{61}
		\left\|\mathbf{\tilde{b}}_1\right\|_2^2=\frac{4\pi^2\cos^2\theta}{ \lambda^2}\mathbf{a}^H\mathbf{D}_1^H\mathbf{\Phi}^H \mathbf{\Phi}\mathbf{D}_1\mathbf{a},
	\end{equation}
	where $\mathbf{a}^{H}\mathbf{D}_1^{H}\mathbf{\Phi}^{H}\mathbf{\Phi}\mathbf{D}_1\mathbf{a}=\|\mathbf{\Phi}\mathbf{D}_1\mathbf{a}\|_2^2$ is expanded as
	\begin{equation}\label{62}
		\begin{aligned}
			&\|\mathbf{\Phi}\mathbf{D}_1\mathbf{a}\|_{2}^{2}   
			=\frac{1}{N}\sum_{k=1}^{K T}\sum_{\dot n\in \Omega(k)}\bar{d}^2_{\dot n}+\\
			&\frac{1}{ N}\sum_{k=1}^{KT}\sum_{\dot n \in \Omega(k)}\sum_{\dot n^{\prime}\neq \dot n}\bar{d}_{\dot n} \bar{d}_{\dot n'}e^{j(\alpha_{k,\dot n}-w_{\dot n})+j(w_{\dot n'}-\alpha_{k,\dot n^{\prime}})},
		\end{aligned}
	\end{equation}
	the $k$-th element of $\mathbf{\Phi}\mathbf{D}_1\mathbf{a}$  is $ \frac{1}{\sqrt{N}} \sum_{\dot n \in \Omega(k)} \bar{d}_{\dot n} e^{j\left(w_{\dot n}-\alpha_{k,\dot n}\right)}$. Similarly, by applying the properties of the characteristic function, the second term is eliminated, and the first term is approximated by $\frac{ T}{N} \sum_{\dot n=1}^{\dot N}\bar{d}^2_{\dot n}$.  Subsequently, $\|\tilde{\mathbf{b}}_1\|_2^2$  is derived that
	\begin{equation}\label{63}
		\left\|\mathbf{\tilde{b}}_1\right\|_2^2\approx \frac{4\pi^2\cos^2\theta  T}{N \lambda^2}\sum_{n=1}^{\dot N}\bar{d}^2_{\dot n},
	\end{equation}
	and $\left| \tilde{\mathbf{b}}_1^H \tilde{\mathbf{a}} \right|^2$ is given by
	\begin{equation}\label{64}
		\left| \tilde{\mathbf{b}}_1^H \tilde{\mathbf{a}} \right|^2=\frac{4\pi^2\cos^2\theta}{\lambda^2}\Big|\mathbf{a}^H\mathbf{D}_1^H\mathbf{\Phi}^\mathrm{H}\mathbf{\Phi}\mathbf{a}\Big|^2,
	\end{equation}
	with $\Big|\mathbf{a}^H\mathbf{D}_1^H\mathbf{\Phi}^H\mathbf{\Phi}\mathbf{a}\Big|^2$ being calculated as:
	\begin{equation}\label{65}
		\begin{aligned}			&\left|\mathbf{a}^H\mathbf{D}_1^H\mathbf{\Phi}^H\mathbf{\Phi}\mathbf{a}\right|^{2}=\frac{1}{N^{2}}\Bigg|\sum_{k=1}^{K T}  \Big(\sum_{\dot n \in \Omega(k)}\bar{d}_{\dot n}e^{j\left(\alpha_{k,\dot n}-w_{\dot n}\right)}\\
			&\sum_{\dot n^{\prime}\in \Omega(k)}e^{j\left(w_{\dot n'}-\alpha_{k,\dot n^{\prime}}\right)}\Big)\Bigg|^{2}\approx\frac{T^2}{N^{2}}\Bigg|\sum_{\dot n=1}^{\dot N}\bar{d}_{\dot n}\Bigg|^{2},
		\end{aligned}
	\end{equation}
	Subsequently, $\left| \tilde{\mathbf{b}}_1^H \tilde{\mathbf{a}} \right|^2$  is derived that
		\begin{equation}\label{66}
		\left| \tilde{\mathbf{b}}_1^H \tilde{\mathbf{a}} \right|^2\approx\frac{4\pi^2\cos^2\theta T^2}{\lambda^2N^2}\Big( \sum_{\dot n=1}^{\dot N}\bar{d}_{\dot n}\Big) ^{2},
	\end{equation}

	Substituting  \eqref{60},  \eqref{61} and   \eqref{66} into   \eqref{55}, we have
	\begin{equation}
		\mathbf{F}_{\pmb{\theta}\pmb{\theta}} \approx  \frac{8N_p T \hat{p}\pi^2\cos^2\theta}{ \sigma_n^2\lambda^2N^2K}\Big(\dot N\sum_{\dot n=1}^{\dot N}{\bar{d}}_{\dot n}^{2}-\Big(\sum_{\dot n=1}^{\dot N}{\bar{d}}_{\dot n} \Big)^2\Big),
	\end{equation}
	
	Similarly, $\mathbf{F}_{\pmb{\phi}\pmb{\phi}}$ is approximately expressed as
	\begin{equation}\label{67}
		\mathbf{F}_{\pmb{\phi}\pmb{\phi}} \approx \frac{8N_p T \hat{p}\pi^2\sin^2\theta}{ \sigma_n^2\lambda^2N^2K}\Big(\dot N\sum_{\dot n=1}^{\dot N}{\tilde{d}}_{\dot n}^{2}-\Big(\sum_{\dot n=1}^{\dot N}{\tilde{d}}_{\dot n} \Big)^2\Big),
	\end{equation}
	
	For $\mathbf{F}_{\pmb{\theta}\pmb{\phi}}$ and $\mathbf{F}_{\pmb{\phi}\pmb{\theta}}$, it is further derived that $\tilde{\mathbf{b}}_1^H \tilde{\mathbf{b}}_2$ is expressed as
	\begin{equation}\label{68}
		\tilde{\mathbf{b}}_1^H \tilde{\mathbf{b}}_2 = \frac{4\pi^2\sin\theta\cos\theta}{\lambda^2}\mathbf{a}^H\mathbf{D}_1^H\mathbf{\Phi}^H\mathbf{\Phi}\mathbf{D}_2\mathbf{a},
	\end{equation}
	with $\mathbf{a}^H\mathbf{D}_1^H\mathbf{\Phi}^H\mathbf{\Phi}\mathbf{D}_2\mathbf{a}$ computed by
	\begin{equation}\label{69}
		\begin{aligned}
			&\mathbf{a}^H\mathbf{D}_1^H\mathbf{\Phi}^H\mathbf{\Phi}\mathbf{D}_2\mathbf{a}  \\
			&= \frac{1}{N} \sum_{k=1}^{KT}\sum_{\dot n\in \Omega(k)} \bar{d}_{\dot n} \tilde{d}_{\dot n}
			+ \\
			&\frac{1}{N} \sum_{k=1}^{K T} \sum_{\dot n \in \Omega(k)} \sum_{\substack{ \dot n^{\prime} \neq \dot n}} \bar{d}_{\dot n} \tilde{d}_{\dot n'}
			e^{j(\alpha_{k,\dot n} - w_{\dot n}) + j(w_{\dot n'} - \alpha_{k,{\dot n}'})} \\
			&\approx\frac{T}{N} \sum_{n=1}^{\dot N} \bar{d}_{\dot n} \tilde{d}_{\dot n},
		\end{aligned}
	\end{equation}

	Next, the term $\tilde{\mathbf{b}}_1^{H}\tilde{\mathbf{a}}\tilde{\mathbf{a}}^{H}\tilde{\mathbf{b}}_2$ is expressed as
	\begin{equation}\label{72}
		\tilde{\mathbf{b}}_1^{H}\tilde{\mathbf{a}}\tilde{\mathbf{a}}^{H}\tilde{\mathbf{b}}_2 = \frac{4\pi^2\sin\theta\cos\theta}{\lambda^2}\mathbf{a}^{H}\mathbf{D}_1^{H}\mathbf{\Phi}^{H}\mathbf{\Phi}
		{\mathbf{a}}{\mathbf{a}}^{H}\mathbf{\Phi}^{H}\mathbf{\Phi}\mathbf{D}_2\mathbf{a},
	\end{equation}
	
	Employing \eqref{65}, $\mathbf{a}^{H}\mathbf{D}_1^{H}\mathbf{\Phi}^{H}\mathbf{\Phi}{\mathbf{a}}{\mathbf{a}}^{H}\mathbf{\Phi}^{H}\mathbf{\Phi}\mathbf{D}_2\mathbf{a}$ is approximately expressed as
	\begin{equation}\label{73}
		\begin{aligned}
			&\mathbf{a}^\mathrm{H}\mathbf{D}_1^\mathrm{H}\mathbf{\Phi}^\mathrm{H}\mathbf{\Phi}{\mathbf{a}}{\mathbf{a}}^\mathrm{H}\mathbf{\Phi}^\mathrm{H}\mathbf{\Phi}\mathbf{D}_2\mathbf{a}\approx\frac{T^2}{N^2}\Big(\sum_{\dot n=1}^{\dot N}\bar{d}_{\dot n}\Big)\Big(\sum_{\dot n=1}^{\dot N}\tilde{d}_{\dot n}\Big),
		\end{aligned}
	\end{equation}
	
	Accordingly, we have
	\begin{equation}\label{74}
		\begin{aligned}
			&\mathbf{F}_{\pmb{\theta}\pmb{\phi}} = \mathbf{F}_{\pmb{\phi}\pmb{\theta}} \\
			&\approx \frac{8N_p T \hat{p}\pi^2\sin\theta\cos\theta}{\sigma_n^2\lambda^2N^2K}\Big(\dot N \sum_{n=1}^{\dot N}{\bar{d}}_{\dot n}{\tilde{d}}_{\dot n}
			- \\ &\Big(\sum_{n=1}^{\dot N}{\bar{d}}_{\dot n}\Big)\Big(\sum_{n=1}^{\dot N}{\tilde{d}}_{\dot n}\Big)\Big),
		\end{aligned}
	\end{equation}
	
	By invoking the (population) variance and covariance operators over the $\dot N$ virtual-element index,  we obtain the following equivalent rewritings:
	\begin{equation}
		\mathbf{F}_{\pmb{\theta}\pmb{\theta}}
		\approx \frac{8N_p \hat{p}\pi^2\cos^2\theta}{ \sigma_n^2\lambda^2}\, T K \,\mathrm{Var}\!\big(\bar d_{\dot n}\big),
	\end{equation}
	
	\begin{equation}
		\mathbf{F}_{\pmb{\phi}\pmb{\phi}}
		\approx \frac{8N_p \hat{p}\pi^2\sin^2\theta}{ \sigma_n^2\lambda^2}\, T K \,\mathrm{Var}\!\big(\tilde d_{\dot n}\big),
	\end{equation}
	
	\begin{equation}
		\mathbf{F}_{\pmb{\theta}\pmb{\phi}}=\mathbf{F}_{\pmb{\phi}\pmb{\theta}}
		\approx \frac{8N_p \hat{p}\pi^2\sin\theta\cos\theta}{ \sigma_n^2\lambda^2}\, T K \,\mathrm{Cov}\!\big(\bar d_{\dot n},\tilde d_{\dot n}\big).
	\end{equation}
	
	% end of Appendix blue color

\end{document}